\documentclass[aoas]{imsart}

\RequirePackage{amsthm,amsmath,amsfonts,amssymb}
\RequirePackage[authoryear]{natbib}
\RequirePackage{graphicx}% uncomment this for including figures
\graphicspath{{figures/}} % Directory in which figures are stored
\RequirePackage{float}
\RequirePackage[colorlinks,citecolor=blue,urlcolor=blue]{hyperref}

\startlocaldefs
\numberwithin{equation}{section}
\DeclareMathOperator*{\E}{\mathbb{E}}

\newcommand{\mK}{\mbox{\textbf{K}}}

\newcommand{\mB}{\mbox{\textbf{B}}}

\newcommand{\vK}{\mbox{\boldmath{$K$}}}
\newcommand{\vtheta}{\mbox{\boldmath{$\theta$}}}
\newcommand{\vbeta}{\mbox{\boldmath{$\beta$}}}
\newcommand{\valpha}{\mbox{\boldmath{$\alpha$}}}

\newcommand{\vgamma}{\mbox{\boldmath{$\gamma$}}}
\newcommand{\vp}{\mbox{\boldmath{$p$}}}
\newcommand{\vY}{\mbox{\boldmath{$Y$}}}

\endlocaldefs

\begin{document}

\begin{frontmatter}
\title{A Statistical Perspective on the Challenges in Molecular Microbial Biology}
%\title{A sample article title with some additional note\thanksref{T1}}
\runtitle{Statistics for Microbiome Data}
%\thankstext{T1}{A sample of additional note to the title.}

\begin{aug}
%%%%%%%%%%%%%%%%%%%%%%%%%%%%%%%%%%%%%%%%%%%%%%
%%Only one address is permitted per author. %%
%%Only division, organization and e-mail is %%
%%included in the address.                  %%
%%Additional information can be included in %%
%%the Acknowledgments section if necessary. %%
%%%%%%%%%%%%%%%%%%%%%%%%%%%%%%%%%%%%%%%%%%%%%%
\author[A]{\fnms{Pratheepa} \snm{Jeganathan}\ead[label=e1]{jeganp1@mcmaster.ca}}
%\author[B]{\fnms{???} \snm{???}\ead[label=e2,mark]{???@???}}
\and
\author[B]{\fnms{Susan} \snm{P. Holmes}\ead[label=e2,mark]{susan@stat.stanford.edu}}
%%%%%%%%%%%%%%%%%%%%%%%%%%%%%%%%%%%%%%%%%%%%%%
%% Addresses                                %%
%%%%%%%%%%%%%%%%%%%%%%%%%%%%%%%%%%%%%%%%%%%%%%
\address[A]{Department of Mathematics and Statistics, McMaster University, \printead{e1}}

\address[B]{Department of Statistics, Stanford University, , \printead{e2}}
\end{aug}

\begin{abstract}
High throughput sequencing (HTS)-based technology enables identifying and quantifying non-culturable microbial organisms in all environments. Microbial sequences have enhanced our understanding of the human microbiome, the soil and plant environment, and the marine environment. All molecular microbial data pose statistical challenges due to contamination sequences from reagents, batch effects, unequal sampling, and undetected taxa. Technical biases and heteroscedasticity have the strongest effects, but different strains across subjects and environments also make direct differential abundance testing unwieldy. We provide an introduction to a few statistical tools that can overcome some of these difficulties and demonstrate those tools on an example. We show how standard statistical methods, such as simple hierarchical mixture and topic models, can facilitate inferences on latent microbial communities. We also review some nonparametric Bayesian approaches that combine visualization and uncertainty quantification. The intersection of molecular microbial biology and statistics is an exciting new venue. Finally, we list some of the important open problems that would benefit from more careful statistical method development
\end{abstract}


\end{frontmatter}
%%%%%%%%%%%%%%%%%%%%%%%%%%%%%%%%%%%%%%%%%%%%%%
%% Please use \tableofcontents for articles %%
%% with 50 pages and more                   %%
%%%%%%%%%%%%%%%%%%%%%%%%%%%%%%%%%%%%%%%%%%%%%%
%\tableofcontents

%%%%%%%%%%%%%%%%%%%%%%%%%%%%%%%%%%%%%%%%%%%%%%
%%%% Main text entry area:
\section{Introduction}

High-throughput sequencing (HTS) enables the characterization of variation in microbial diversity in naturally changing or experimentally perturbed environments. Several technologies in molecular microbiology have revolutionized the resolution at which many environments can now be studied. The first is marker-gene sequencing (usually identified as microbiome studies); these use small regions (V4-V6) of a particular gene (the 16S rRNA gene, occasionally others) that serves as a ``fingerprint" or signature for each species or strain of bacteria. More comprehensive profiling of all genes present in these microbial communities is available through complete shotgun sequencing. Shotgun sequence analysis, known as metagenomics, uses all the nucleic acids in a specimen to provide a comprehensive inventory of both the genes and taxa present, sometimes invoking what is known as the \textit{pangenome} \citep{Quince2017desman}. This can be done by Bayesian classifiers using short strings of nucleotide (k-mer) occurrences \citep{Rosen2011} or by genome assembly \citep{lu2017bracken}. \cite{Quince2017shotgun} presents a review and comparison of these different approaches.

These two basic sequence-based methods have enabled major advances in biological, agricultural, and environmental research. For example, vaginal microbiome studies in pregnant women have shown that the reduced \textit{Lactobacillus} species in the vagina is a risk factor for premature birth \citep{callahan2017, digiulio2015}. The gut microbiome is associated with an increased risk of type 1 diabetes and inflammatory bowel diseases in children \citep{kostic2015dynamics, gevers2014treatment}. Outside of human biology, microbial biota has been an important source for monitoring the marine environment \citep{thompson2017, gilbert2014}. Recent work in soil science has also shown the power of cross-sectional microbiome experimental designs to detect ecological perturbations \citep{delgado2016microbial, ramirez2018detecting} and enable climate change monitoring \citep{cavicchioli2019scientists}. A recent review of microbiome-based agro-management has shown how these can improve agricultural production, promote plant growth and health, maintain resistance against diseases, and quantify environmental stress \citep{compant2019}. 

The statistical analyses of abundance and diversity of microbial sequence data have many commonalities with standard ecological studies; this means that many of the downstream tools are already available in the statistical ecologists' toolbox. Spatial, multivariate, and longitudinal methods are central to the field and through the development of tools such as {\tt phyloseq} \citep{mcmurdie2013} we have tried to create bridges between the raw molecular genomic read data (as well as the phylogenetic relationships between taxa) and the data structures such as contingency tables already used in ecology and evolutionary biology.

The current review will concentrate on the statistical challenges inherent in the analyses of the sequencing reads organized as contingency tables. We will use the columns to represent the biological specimens, also called ``specimen-samples," as in Table \ref{tab:1}, the rows are labeled for the taxonomic strains known as Amplicon Sequence Variants (ASVs) \citep{callahan2017exact}. These rows are not predefined before the data become available and are inferred by denoising the raw sequences using the read frequencies and their quality scores, see \cite{callahan2016, DADA2}.
Contrary to some recent statements in the literature \citep{quinn2018understanding, gloor2017microbiome, silverman2017phylogenetic}, the data themselves cannot be considered \textit{compositional} since the number of rows (i.e., strains) and their definition is not known a \textit{priori}, and there is always a substantial and variable proportion of reads that cannot be annotated. The total number of reads in each column corresponds to sequencing depths for each of the specimens and are often called the library sizes: we will show that these can be modeled as gamma-Poisson random variables. Strain level resolution is now also available for shotgun metagenomic data through Bayesian co-occurrence analyses (\cite{Quince2020metagenomics}), and many new research problems arise when this higher resolution of analysis is used.

At high resolution, the unknown parameters of interest are the true prevalence of each of the microbial strains or ASVs and the differences between prevalences across different treatment groups or locations. The strain prevalence parameters sum to one within each specimen $j$, thus if we restrict ourselves to a finite number $(m)$ of possible taxa, the $\vp_{j}=(p_{i,j}, i=1, \dots ,m)$ {\bf do} belong to the simplex. However, even if we postulate \textit{a priori} knowledge on $m$, the estimation of the ${\mathbf p_{j}}$ parameters poses several challenges. There is often between subject or location variability in the different strains detected in the specimens; thus, strains do not co-occur systematically across subjects or locations, and the count table is sparse with a large proportion of zeros (both technological and biological). There are also substantive experimental issues with DNA extraction and PCR amplification that preclude direct quantification of prevalences \citep{mclaren2019consistent}.

Many authors start their analysis by transforming the reads to relative abundances (transforming the counts by dividing by column sums) and then taking the log \citep{Kurtz2015}. However, all versions of centered log-ratio transformations ignore the underlying heteroscedasticities of the prevalence estimates due to library size variation and hinder valid downstream statistical inference. A statistical solution we recommend is variance stabilizing transforms similar to those applied to other genomic data such as RNA-seq, see  \citep{mcmurdie2014}. If there is a small number of pre-specified taxa measured, it is possible to use weighted chi-square distances to account for compositionality, as proposed for ecological tables by \cite{greenacre2010log, greenacre2011compositional}, this incorporates the column sums and thus does not ignore any of the information in the data. In cases where the rare taxa are the study focus, it can be beneficial to transform the data into a 0/1 table with just indicators of presence for taxa that appear abundant above a certain number of reads (typically 2 to 4, depending on the library sizes). On the other hand, if comparisons are to be made between abundances in the core microbiome (a set of common taxa present in most of the specimens), the rare species will be filtered out. As is the case with all statistical transformations, the choice has to be an informed one, and there is no one method that fits all situations, as some environments are infinitely diverse, whereas others are very sparse. Starting by doing  the robust rank-threshold transformation we describe 
in Section \ref{trans} is often a good first step. 

Currently, many experiments that involve microbiota also include complementary assays that provide metabolic and gene expression profiles through the use of Mass Spectrometry \citep{MassSpecGregory2018} and RNA-seq transcriptomics \citep{franzosa2014relating}. These data also include clinical or chemical covariates measured on the same specimens facilitating a ``holistic" understanding of the system under study. A recent review by \cite{sankaran2019multitable} shows how many data integration approaches based on matrix decompositions and cross-table correlations can combine multiple data domain types effectively. We will not cover the multidomain-multimodal aspects here. 

In Section \ref{data}, we introduce the form in which the data are collected, the format of the tables and contiguous information, and the statistical notation. Then, in Section \ref{exa}, we describe an example problem and the data we will use as an illustration of the different tools available. Goodness of fit tests for each taxon enabled us to build generative models for these data that we can use for simulation or power studies such as the one illustrating the strain switching problem in Section \ref{permanova}. We also recommend  Bayesian hierarchical models that we used to remove DNA contamination\textemdash a common additive source of measurement error. For the downstream analysis, it is first necessary to account for the library size - a source of multiplicative variation - discussed in Section \ref{models}. We then describe best practices for variance stabilization and robust rank-based transformation, as well as first-order rank-based dimensionality reductions in Section \ref{trans}. We recommend starting all analyses with data exploration and visualizations of both the raw and transformed count data, including heatmaps and clustering of the specimens illustrated in Section \ref{vis}. We discuss network analysis, many combinations of different distances, multivariate methods, and how we can enhance ordination using phylogenetic trees that account for evolutionary relationships in the taxa in Section \ref{multi}. In each of the above sections, we show how these exploratory tools may not identify high-resolution variability and may be challenging to interpret.  

An important goal in microbial ecology is the inference of differences in taxonomic abundances in different environments or treatment groups. In Section \ref{diff}, we briefly review
permutational analyses of distance matrices, 
generalized linear models for differential abundance methods and discuss the importance of identifying the bacterial communities and their differences across conditions. This motivates the use of latent Dirichlet allocation (LDA) topic models that we discuss and illustrate in Section \ref{LDA}.  To help the biologists interpret the topics, we  enhance this topic model analysis with a visualization that incorporates the phylogenetic tree, and we show it in the same section. We identify promising research direction is Bayesian nonparametric approaches that can directly account for the uncertainty in measured data, learning latent structure, and flexible enough through the use of hierarchies that can account for experimental design and random effects. For example, to account for the growing number of taxa and uncertainties in ordination analyses, we demonstrate a Bayesian nonparametric factor ordination method in Section \ref{BN}. We conclude by discussing the example dataset results and open problems from a statistical perspective.

\section{Microbiome data} \label{data}
\begin{figure}[H]
	\centering
	\includegraphics[width=\linewidth]{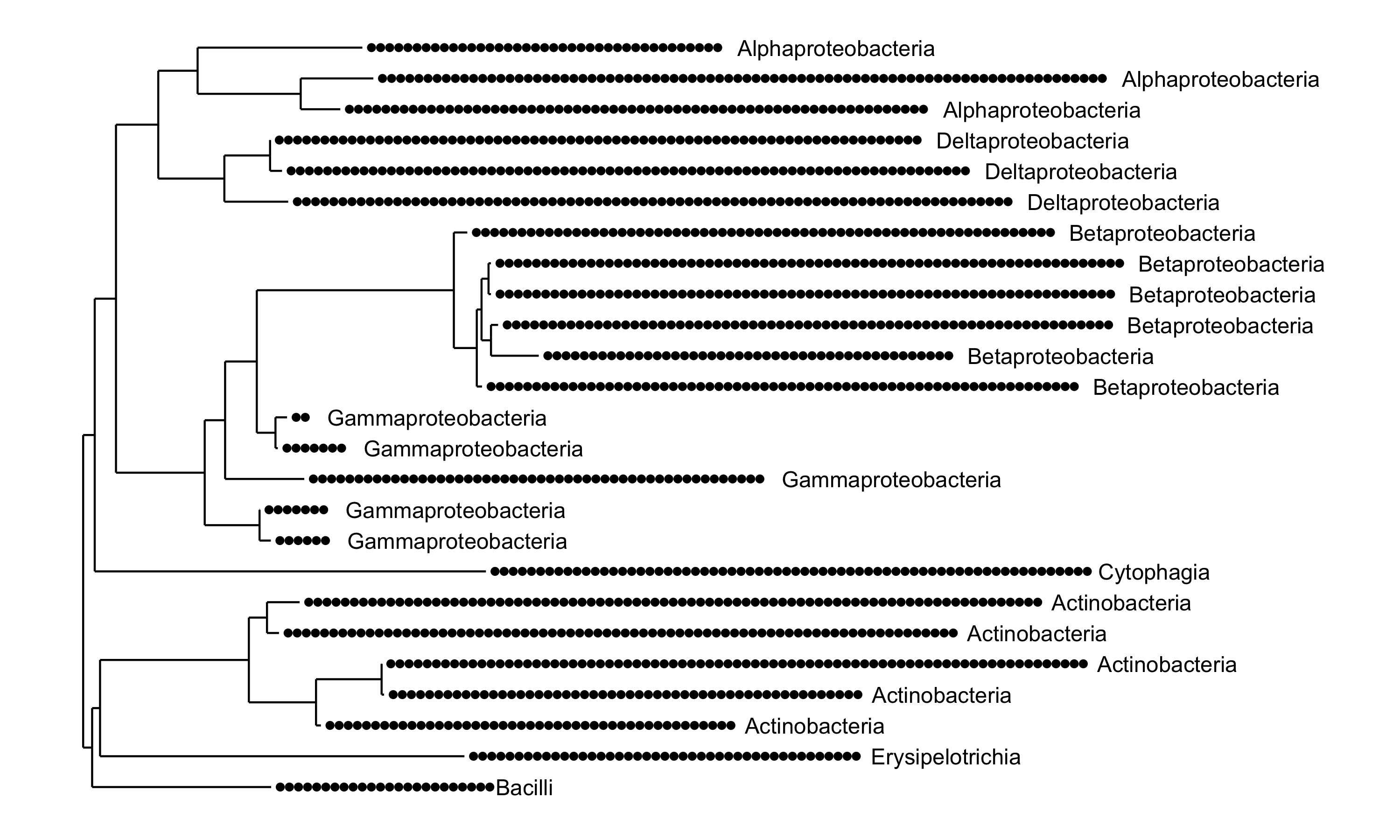}
	\caption{Phylogenetic tree. Black points at each node corresponds to the specimens in which the corresponding taxa is present. Tip labels are class of the taxon.}
	\label{fig01}
\end{figure}

In both marker-gene sequencing and shotgun metagenomics, the core of the data consists of a contingency table of read counts with specimens recorded in the columns and taxa (ASVs) identified in the rows as in Table \ref{tab01}. We will use marker-genes throughout the text, but the methods and challenges discussed also apply to shotgun metagenomics. Associated with these counts is a matrix of specimen information with specimen identifiers on rows and column variables such as subject identity, sequencing batch, type of specimen (control or specimen types such as blood or gut or placenta, etc.), as shown in Table \ref{tab02}. It is often beneficial to annotate the Amplicon Sequence Variants (ASVs) \citep{callahan2017exact} using a matrix of taxa identifiers at several selected taxonomy levels such as species name on rows and taxonomy levels (species, genus, phylum, family, order, class) on columns as shown in Table \ref{tab03}. Finally, evolutionarily relationship between taxa are formalized as a phylogenetic tree as in Figure \ref{fig01}.

\begin{table}[H]
	\caption{\label{tab01}Count matrix $\mK \in \mathbb{R}^{m \times N}$ of $n_{1}$ specimens, $n_{2}$ controls and $m$ taxa, where $K_{ij}$ are the reads of taxon $i$ in $j$-th specimen and $K_{il}^{0}$ denote the number of reads of taxon $i$ in $l$-th negative control.}
	\centering
	\begin{tabular}{|c| c c c c| c c c c|}
		\hline
		& $\text{Specimen}_{1}$ & $\text{Specimen}_{2}$ &$ \cdots $&$\text{Specimen}_{n_{1}}$&$\text{Control}_{(n_{1}+1)}$&$\text{Control}_{(n_{1}+2)}$&$\cdots$ & $\text{Control}_{N}$\\
		\hline
		\hline
		$\text{Taxa}_{1}$& $K_{11}$ & $K_{12}$ & $ \cdots  $ & $K_{1n_{1}}$ & $K^{0}_{1(n_{1}+1)}$ & $K^{0}_{1(n_{1}+2)}$ &$\cdots$ &$K^{0}_{1N}$\\
		
		$\text{Taxa}_{2}$& $K_{21}$ & $K_{22}$ & $ \cdots $ & $K_{2n_{1}}$ & $K^{0}_{2(n_{1}+1)}$ & $K^{0}_{2(n_{1}+2)}$ &$\cdots $ &$K^{0}_{2N}$\\
		
		$	\vdots$ &$\vdots$ & & & & & $\vdots$ & &$\vdots$ \\
		
		$\text{Taxa}_{i}$& $K_{i1}$ & $K_{i2}$ & $ \cdots  $ & $K_{in_{1}}$ & $K^{0}_{i(n_{1}+1)}$ & $K^{0}_{i(n_{1}+2)}$ &$\cdots$ &$K^{0}_{iN}$\\
		
		$	\vdots$ &$\vdots$ & & & & & $\vdots$ & &$\vdots$ \\
		
		$\text{Taxa}_{m}$& $K_{m1}$ & $K_{m2}$ & $  \cdots $ & $K_{mn_{1}}$ & $K^{0}_{m(n_{1}+1)}$ & $K^{0}_{m(n_{1}+2)}$ &$\cdots $ &$K^{0}_{mN}$\\
		\hline
	\end{tabular}
	\label{tab:1}	
\end{table}

\begin{table}[H]
	\caption{\label{tab02}Sample data, a matrix of specimen information with specimen identifiers on rows and column variables. }
	\centering
	\fbox{%
		\begin{tabular}{c |c c c c }
			& Specimen ID & Subject ID & Specimen type& Batch number \\
			\hline
			$\text{Specimen}_{1}$ & $\text{Specimen}_{1}$ & $\text{Subject}_{1}$ & Plasma&1\\
			$\text{Specimen}_{2}$ & $\text{Specimen}_{2}$ & $\text{Subject}_{2}$ & Plasma&2\\
			$\vdots$ &  $\vdots$ & $\vdots$ & $\vdots$ & $\vdots$ \\
			$\text{Specimen}_{n_{1}}$ & $\text{Specimen}_{n_{1}}$ & $\text{Subject}_{n_{1}}$ & Plasma&1\\
			
			$\text{Control}_{(n_{1}+1)}$ & $\text{Control}_{(n_{1}+1)}$ & $\text{Reagent}$& Control&1\\
			
			$\text{Control}_{(n_{1}+2)}$ & 	$\text{Control}_{(n_{1}+2)}$ &
			$\text{Library}$& Control&1\\
			
			$\vdots$ & $\vdots$ & $\vdots$ & $\vdots$ &$ \vdots$ \\
			$\text{Control}_{N}$ & $\text{Control}_{N}$ & $\text{Reagent}$& Control&2
	\end{tabular}}
\end{table}

\begin{table}[H]
	\setlength{\tabcolsep}{.3pt}
	\caption{\label{tab03}Taxonomy table }
	\begin{tabular}{|c| c| c |c |c |c| c| }
		\hline
		& Kingdom & Phylum& Class &Order& Family & Genus \\
		\hline
		$\text{Taxa}_{1}$& \textit{Bacteria} & \textit{Nitrospirae} & \textit{Nitrospira}& \textit{Nitrospirales}&\textit{0319-6A21}  &\\
		
		$\text{Taxa}_{2}$& \textit{Bacteria} & \textit{Acidobacteria} & \textit{	Blastocatellia}& \textit{Blastocatellales}&\textit{Blastocatellaceae\_(SG\_4)}  &\textit{DS-100} \\
		
		$	\vdots$ &$\vdots$ & & &  & & \\
		
		$\text{Taxa}_{i}$& \textit{Bacteria} & \textit{Armatimonadetes	} & \textit{Armatimonadia}& \textit{Armatimonadales}&\textit{Armatimonadaceae}  &\textit{Armatimonas} \\
		
		$	\vdots$ &$\vdots$ & & &  & & \\
		
		$\text{Taxa}_{m}$& \textit{Bacteria} & \textit{Chloroflexi} & \textit{Chloroflexia}& \textit{Herpetosiphonales}&\textit{Herpetosiphonaceae}  &\textit{Herpetosiphon}  \\
		\hline
	\end{tabular}
\end{table}

Table \ref{tab01} shows the count table of $n_{1}$ specimens, $n_{2}$ controls and $m$ taxa, where $K_{ij}$ are the reads of taxon $i$ in $j$-th specimen and $K_{il}^{0}$ denote the number of reads of taxon $i$ in $l$-th negative control - specimens consisting of molecular-grade water included in each extraction batch. We denote $d_{j}$ and $d_{l}^{0}$ the linear scaling factors for specimen $j$ and control $l$ that account for the library sizes --- sums of the columns in the count Table \ref{tab01} and can vary by orders of magnitude. For each specimen $j$, the read counts are in the vector $\vK_{j}=\begin{bmatrix} K_{1j}, K_{2j}, \cdots, K_{mj} \end{bmatrix}^{T}.$  Note that many bacteria are interdependent (in particular syntropic relations are frequent) precluding the use of a simple multinomial for these counts. We provide the details of a typical analysis for a real example in what follows.

\section{Example of 16S \lowercase{r}RNA gene sequencing data}\label{exa}
We reanalyzed the 16S rRNA gene sequencing of the root endosphere specimens from \cite{fitzpatrick2018}. We retrieved this dataset from Google Dataset Search with the keyword of 16S rRNA gene sequencing and ASV. The authors described the specimen collection, library preparation and sequencing, and their microbial bioinformatics workflow. In plants, host organellar sequences such as mitochondrial and plastid share similar bacterial lineages and reduce the efficiency of quantifying microbial sequences. Universal peptide nucleic acids (PNA) have been used to limit the amplification of host-derived mitochondrial and plastid sequences. \cite{fitzpatrick2018} designed their experiment to evaluate the efficacy of universal plastid peptide nucleic acids (O-pPNA) and Asteraceae-modified pPNA (M-pPNA) in limiting plastid host contaminants. Contamination varies across host plants, and \cite{fitzpatrick2018} provided a validated framework for Asteraceae-modified pPNA so that there is no effect of pPNA type on bacterial detection. \cite{fitzpatrick2018} identified less plastid contaminant sequences in Asteraceae with M-pPNA than O-pPNA. We used this data set as an example that shows typical analytical challenges posed by such data and provided some solutions inspired by statistical approaches. 

We identified the host contaminant sequences from the taxonomy table. The data set had specimens from six Asteraceae and three non-Asteraceae plant types. The O-pPNA applied to three to ten replicates, whereas M-pPNA applied to three to five replicates. The root endosphere specimens from four out of six Asteraceae and all three non-Asteraceae plant types were sequenced with both pPNA (O and M) and had 18 paired specimens as in Supplementary Table \ref{tab04_paired}. A common environment was chosen to grow all plant types from sterilized seeds, except \textit{Centaurea solstitiali}, which were sampled in a field across France, Spain, and the USA. 

We followed the filtering in \cite{fitzpatrick2018}. From 57,116 ASVs, we removed ASVs that lacked a kingdom assignment or were assigned to Archaea or Eukaryota, ASVs that lacked a phylum assignment, ASVs classified as mitochondria, ASVs classified as plastids. We retained plant types that had both endosphere or rhizosphere specimens and negative controls, removed specimens with less than 800 reads, and removed the fifth sequencing run that was bad. In the preprocessed data, there were 86 root endosphere and 25 control specimens for further analysis. The 6929 ASVs in specimens and controls were used in the Bayesian hierarchical model as implemented in the {\tt BARBI} package to remove contaminants \citep{cheng2019combined}.

\section{Models}\label{models}

Although many analyses begin with exploratory analysis of transformed count data that identify outliers,  biological variability and the interrelation between four different components of microbiome data discussed in Section \ref{data}, they cannot account for heterogeneity in the data. It can be beneficial to start by a simple generative model for the individual taxonomic counts. We start by describing a negative binomial (gamma-Poisson) goodness of fit test for each taxon. The test results justify this model and we will complement them with a Bayesian hierarchical model that removes DNA contamination, a supplementary additive error. 

\subsection{Goodness of fit  for taxon counts}
If we knew the true prevalence of different taxa $p_{ij}$ and then sequence the same amplified DNA in technical replicates,  we would expect to see a simple  $\text{Poisson}(\lambda_{ij})$  variation in the $K_{ij}.$  For instance, \cite{grumaz2016} uses a Poisson model with healthy controls providing baseline proportions of each taxon to identify bacteria in blood infected patients after removing the well-known contaminant species \textit{Xanthomonas}. A Poisson model with intensity estimated using negative and positive controls was used in \cite{hong2018} to choose taxa for downstream analysis. However, this simple model needs to be enriched to acomodate other sources of variation, such as contaminant bacteria introduced during the sample preparation, sequencing run batch effects, and library size differences as well the essential biological variation of interest.

We show that a negative binomial distribution  (or equivalently gamma-Poisson) fits our example data set for the ASV counts well. We test the null hypothesis, $H_{0}$,  that the ASV counts have a negative binomial distribution using a chi-square test statistic. We draw 1000 simulations from the negative binomial with the parameters estimated from the data (see the supplementary Rmd and html files in github at \href{https://pratheepaj.github.io/diffTop/ }{https://pratheepaj.github.io/diffTop/ }).  We can then compute the p-value for the observed ASVs. The Supplementary Figure \ref{fig02} shows how uniformly distributed the p-values are under the null hypothesis. After controlling for multiple testing, no ASVs reject the negative binomial distribution. For some ASVs (2.3\% of total ASVs), the presence of zeros are larger than expected under negative binomial distribution. Some researchers have preferred zero-inflated negative binomial distribution (see \cite[Chapter 4]{holmes2018modern}  for a definition and formula) for such data \cite{xu2015assessment}. Comparing the two models can be done using the Akaike Information Criterion (AIC) as in \cite{romero2014vaginal}.  

In the case of shotgun metagenomics data, it is also necessary to add  taxonomic ``bias" factors \citep{mclaren2019consistent}. Several of these unknown parameters have multiplicative effects, whereas the sequencing count data are also influenced by additive noise, such as contaminating DNA from reagents and the environment. Finally, we note that the sum of independent negative binomial random variables with the same parameters also follows a negative binomial distribution, so the library sizes are also expected to be negative binomial.

\subsection{Additive and Multiplicative  Errors}
Specimens that are sequenced at much smaller library sizes will show many more zeros. Thus, zero-inflation is correlated with specimen library size and can be included in the relevant mixture models. In the zero-inflation case, the data can be split into the core taxa that present few zeros and whose abundances are used and a presence-absence table that encodes presence at a minimum number of reads and can take the rarer taxa into account.

Contamination of sequence-based data is modeled as an additive error \citep{davis2018, salter2014}. For instance, \cite{salter2014} show reagents used in DNA extraction kits are heavily contaminated with microbial DNA, resulting in background noise that critically impacts results. \cite{davis2018} proposed a simple statistical method called {\tt decontam}, that identifies DNA contaminants in microbial studies using prevalences in designed experiments where negative controls have been included or by using frequencies if DNA concentrations of each specimen are included in the sample data Table \ref{tab02}. However, {\tt decontam} may not identify specimen specific DNA contamination or rare taxa.

For contaminant removal, in the presence of negative controls, it has been shown that a statistical mixture model can estimate each taxon's true intensity using reference priors and enable the removal of contaminant taxa \citep{cheng2019combined}. The Bayesian hierarchical model for inferring DNA contamination in each specimen is as follows. 
\begin{equation}
\label{eq:hieModel}
\begin{split}
K_{ij} \vert \lambda_{ij}^{(r)},\lambda_{ij}^{(c)}, d_{j} &\sim \text{Poisson} \left(\left(\lambda_{ij}^{(r)} + \lambda_{ij}^{(c)}\right)d_{j}\right),\\
\lambda_{ij}^{(r)}\sim p \left(\lambda_{ij}^{(r)} \right) &= \dfrac{\left|  I\left( \lambda_{ij}^{(r)} \right) \right|^{1/2}}{\left|  I\left(0 \right) \right|^{1/2}}, \hspace{.3in}
\lambda_{ij}^{(c)} \sim \text{gamma}\left(\alpha_{ij}^{(c)}, \beta_{ij}^{(c)}\right),
\end{split}
\end{equation}
where $\lambda_{ij}^{(r)}$ is the the true intensity parameter, $\lambda_{ij}^{(c)}$ is the contaminant intensity parameter, $p \left(\lambda_{ij}^{(r)} \right)$ is a marginal reference prior for the true intensities.  We define it as a function of the Fisher information obtained through the marginal probability densities of $\lambda_{ij}^{(r)}$, $I\left(\cdot\right) = -\E \left[\left( \dfrac{\partial^{2}}{\partial (\lambda_{ij}^{(r)})^{2}} \log p\left(k_{ij}\vert  \lambda_{ij}^{(r)}, d_{j} \right)\right) \right].$ We estimate hyper parameters $\alpha_{ij}^{(c)}$ and $\beta_{ij}^{(c)}$ using negative controls and find the reference for the library size using the median of ratios method \citep{anders2010} (see Section \ref{libSc}). We construct 95\% highest posterior density (HPD) interval for the true intensity $\left(L_{ij}^{(r)}, U_{ij}^{(r)}\right)$ and the contaminant intensity $\left(L_{ij}^{(c)}, U_{ij}^{(c)}\right)$ for each taxon $i$ in a specimen $j$. We declare a taxon to be contaminant taxon if the lower limit $L_{ij}^{(r)}$ is smaller than the upper limit $U_{ij}^{(c)}.$ For the 6,929 ASVs in specimens and controls, we applied {\tt BARBI}  \citep{cheng2019combined} and detected and removed 1,121 contaminants ASVs. We used the remaining 5,808 ASVs in 86 specimens for our downstream analysis.

\subsection{Library size scaling factor}\label{libSc}

All downstream analyses need to account for  library sizes (the column sums of the ASV contingency table); they are random multiplicative factors. In the context of RNA-seq, \cite{anders2010} propose a median-of-ratios algorithm that works well to estimate a library size scaling factor for each specimen. This method divides each taxon count in Table \ref{tab01} by the row's geometric mean, and the library size scaling factor $d_{j}$ is the median of the ratios for each specimen $j$. After the library size adjustment by $d_{j}$, an appropriate variance stabilization is applied and illustrated  in Section \ref{trans}.

Here we review some of the transformations we have found useful for the microbial count table. These transformations reduce the systematic variation in the microbiome count table and make the data approximately homoscedastic. Then we use dimensionality reduction methods to explore possible hierarchy factors or any batch effects.

\section{Transformations}\label{trans}

\subsection{Variance stabilization}

Several parametric and nonparametric transformations were proposed for  microbial count  contingency tables in \cite{mcmurdie2014}. The variance of variance-stabilized transformed value is approximately independent of its mean value. Nonparametric regression methods are often used to characterize the mean-variance dependence of the library size normalized data. To do this, we can compute the mean and variance for each taxon. We then use a nonparametric regression method such as LOESS to model the relationship between mean and variance: the weights of each observation are adjusted using this fit, and an inverse transformation is applied that stabilizes the variance. 

The transformations provided in {\tt Voom} \citep{law2014voom} scale the count table to count per million (CPM), then use the nonparametric fit for mean-variance dependence to compute the weight for each observation, take the log transformation of weighted observation to obtain variance-stabilized values. \cite{Fukuyama2017-c} shows that CPM transformations tend to produce false positives at the differential abundance analysis step. In our case, where the counts follow a negative binomial (NB) distribution, the \cite{anscombe1948transformation} transformation provides an analytical solution. Given $\vK_{j}=\begin{bmatrix} K_{1j}, K_{2j}, \cdots, K_{mj} \end{bmatrix}^{T}$, we let $K_{ij}$ be a draw from a NB distribution with mean $\mu_{i}$ and dispersion/exponent $k_{i}$. \cite{anscombe1948transformation} shows that the optimal transformation for NB distribution is $K^{*}_{ij}$, where $K^{*}_{ij}= \text{sinh}^{-1} \left(\sqrt{\dfrac{K_{ij}+c}{k_{i}-2c}} \right)$. For $k >2$ and $\mu_{i}$ large, $c = \dfrac{3}{8}$ and $\text{var}\left(K^{*}_{ij} \right) = \dfrac{1}{4}\psi^{'}\left(k_{i}\right),$ where $\psi^{'}\left(k_{i}\right) = \dfrac{1}{k_{i}-1/2}$ for large $k_{i}$. All these types of transformations are now available in DESeq2 \citep{love2014}. We propose \cite{anscombe1948transformation} transformation to stabilize the variance (see Figure \ref{fig04} in the Supplement, as compared to Figures \ref{fig06},  \ref{fig07}, and  \ref{fig08}).

\subsection{Rank-based methods}
A robust approach to testing for treatment effects is to rank the taxa within each specimen from the most frequent to the least frequent. Most specimens do not contain representatives from all taxa, and noise level read counts could create large jumps in ranks at the lower end. Thus a threshold-rank approach is preferable, where the lower ranks are all assigned the same tied value of one. For instance, suppose there are 1,000 taxa present in the full data, and about one third occur at noise level in several specimens. The rank of noisy taxa could jump from 1 to 330 just by chance. Thus we assign scores from 670 for the most frequent in a specimen, down to a tied score of one for the last 330 taxa.

To choose the threshold 330 above, we performed a preliminary study on presence-absence patterns and choose the threshold for the number of reads as indicators that a taxon is present. Typically, three reads are sufficient. We then created a new table with a presence denoted by one and absence zero, $B_{ij}=1$, if $K_{ij} \geq 2.$ We can also represent the binary indicator $B$ as a bipartite graph and rearrange the rows to reveal eventual block structures indicating communities stochastic block models can be convenient tools for such an approach \citep{snijders1997estimation}.

\begin{figure}[H]
	\centering
	\includegraphics[width=\linewidth]{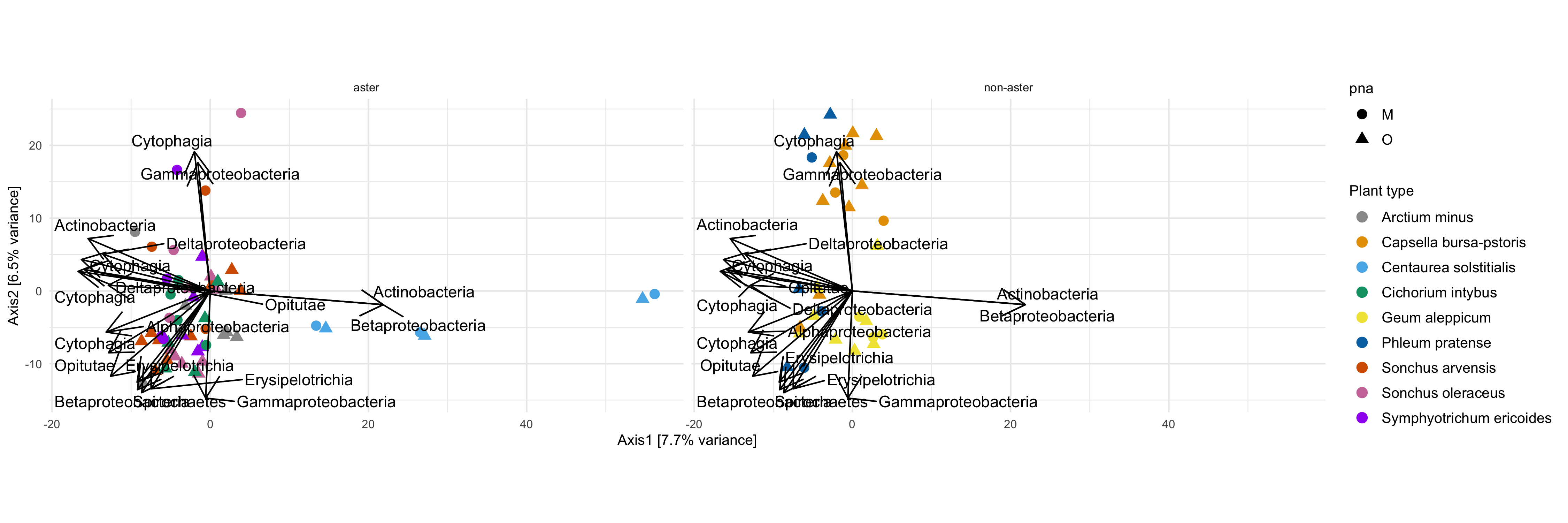}
	\caption{The biplot from the PCA after the truncated-rank transformation. The shape denotes universal (O) and Asteraceae-modified (M) pPNA types, respectively. Facet denotes Asteraceae or non-Asteraceae plants. {\it Actinobacteria} and {\it Betaproteobacteria} dominate in outliers {\it Centaurea solstitialis} in the positive direction of the first axis. The second axis explains the plant microbiome variability in non-Asteraceae specimens.}
	\label{fig3Vis}
\end{figure}

Principal component analysis (PCA) on truncated-rank transformed abundances produces a robust dimension reduction in the presence of a heavy-tailed distribution of count table. For the example data set, we choose ASVs in at least two specimens with at least 25 reads. This filtering results in 1418 ASVs in 86 specimens. Figure \ref{fig3Vis} shows a biplot resulting from the PCA after the truncated-rank transformation. {\it Actinobacteria} and {\it Betaproteobacteria} dominate in outlier {\it Centaurea solstitialis} in the positive direction of the first axis. The second axis explains the plant microbiome variability in non-Asteraceae specimens. The paired-specimens in Supplementary Table \ref{tab04_paired} are relatively close to each other.

\section{Visualizations for heterogeneous data}\label{vis}

Microbiome data are high-dimensional and our example data have the four different components (count matrix, sample data, taxonomic table, phylogenetic tree) as elaborated in Section \ref{data}. Visualization methods identify taxonomic patterns or differences in various phenotypes or temporal variations in longitudinal experiments. These methods are also useful in getting insights from statistical inferences, such as differential abundance analysis. The {\tt phyloseq} \citep{mcmurdie2013} package incorporated wrappers for making interactive layered plots for microbiome data using the {\tt ggplot2} \citep{wickham2016ggplot2} package. 

Traditionally, scientists prefer to use bar plots of the estimates of $\vp_{j}=(p_{i,j}, i=1\ldots m)$. More recently, heatmaps and interactive visualization of the phylogenetic tree jointly with taxonomic frequencies along longitudinal axes such as that provided by {\tt treelapse} \citep{sankaran2018interactive,treelapse} have become popular (see \cite{kuntal2019visual} for a review.) Networks and trees are useful aids to interpretation for microbial communities. Taxa co-occurrence graphs can serve as the basis for nonparametric tests inspired by the Friedman-Rafsky approach. One implementation is the {\tt phyloseqGraphtest} package (\cite{Fukuyama2020} illustrated in \cite[Section ]{callahan2016}).

Simple statistical summaries show that in the example data set with 1418 ASVs in 86 samples, \textit{Proteobacteria} is the most prevalent Phylum. Next largest prevalent Phyla are \textit{Actinobacteria}, \textit{Bacteroidetes}, and \textit{Firmicutes}. We computed the relative abundance of ASVs in each specimen. Then, we removed Class with less than five ASVs. Supplementary Figures \ref{fig09} and \ref{fig10} show the distribution of the relative abundance of each Phylum in specimens faceted by Class. We can use the bar plots to compare the difference in scale and distribution of Phylum in both pPNA types. These figures show unimodal abundance profiles in four different Classes of \textit{Proteobacteria}. There are few replicates sequenced with M-pPNA than O-pPNA in non-Asteraceae. 

\begin{figure}[H]
	\centering
	\includegraphics[width = \linewidth]{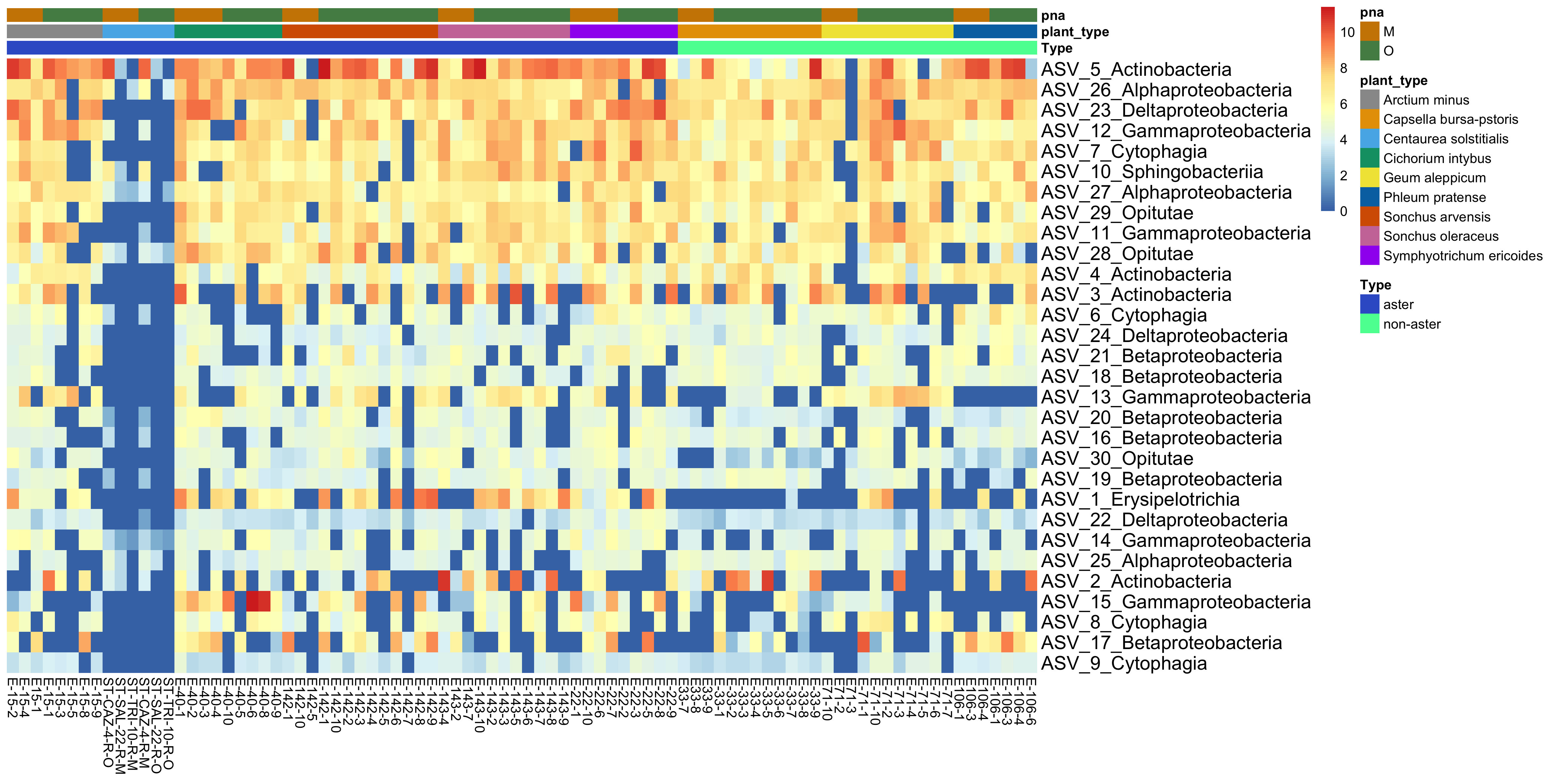}
	\caption{Thirty most abundant ASVs were selected in all specimens. Taxa are labeled by Class on rows, and specimens are on the columns of the heatmap. Some specimens from \textit{Centaurea solstitialis} plant have the most abundant ASVs of the Class \textit{Actinobacteria}, \textit{Alphaproteobacteria}, \textit{Sphingobacteria}, and \textit{Gammaproteobacteria}.}
	\label{fig11}
\end{figure}

The heatmap is constructed using transformed data to maximize the contrasts and understand how the specimens cluster and batch effects. In the example data set, Figure \ref{fig11} shows that the ten most prevalent Class taxonomy is similar in Asteraceae and non-Asteraceae plants, except in \textit{Centaurea solstitialis} plants, which have the most abundant ASVs of the Class \textit{Actinobacteria}, \textit{Alphaproteobacteria}, \textit{Sphingobacteria}, and \textit{Gammaproteobacteria}. The most prevalent Class {\it Actinobacteria} is more abundant in Asteraceae than it is non-Asteraceae plants. 

\section{Multivariate and Network Analyses}\label{multi}

\subsection{Ordination}

To identify outliers, clusters, and relative position or gradients of specimens in low dimensions,  ordination methods can be used  on any of the many distances available for measuring similarities using abundances or presence-absence. Historically, ecologists have been leaders in their careful choice of distances between specimens. The most popular include the Bray-Curtis, chi-square, Wasserstein, or Jaccard distances. Jaccard, Bray-Curtis, or unifrac distances are all popular choices in microbiome studies.  The distance matrices are then used to create sample maps through ordination methods such as multidimensional scaling (MDS) (also known as principal coordinate analysis (PCoA)), double principal coordinate analysis (DPCoA), or nonmetric multidimensional scaling (NMDS). The resulting two or three dimensional maps can indicate clusters of samples when there is an underlying latent categorical variable \citep{mcmurdie2013} or gradients --- continuous latent variables such as water depth seen in the TARA ocean data \citep{nguyen2017bayesian}. Clusters can lead to a simplification of the data by assigning specimens a state type, as is done for the vaginal microbiome studies \citep{romero2014composition,digiulio2015}. However, sometimes clusters appear through artifacts, such as low-density sampling of the high dimensional space \citep{gorvitovskaia2016interpreting}. \cite{nguyen2019ten} provide a set of guidelines to use and interpret dimensionality reduction methods for different data types. 

Supplementary Figure \ref{mds_unifrac} shows an MDS plot built using weighted unifrac distances between all our specimens. Paired specimens are labeled, and we can detect some of them form clusters, except paired-specimens (E-143-7, E143-7), (E-142-5, E142-5), (E-71-2, E71-2), and (E-71-3, E71-3). {\it Centaurea solstitialis} specimens are outliers among Asteraceae plants in the positive direction of Axis 2, which were sampled in a field across three countries. Axis 1 explains the microbial variability in plant types. Supplementary Figure \ref{fig1Vis} shows biplots built using the double principal coordinate method (DPCoA) that also incorporates the phylogenetic information into the ordination. The labels in Supplementary Figure \ref{fig1Vis} (A) depict paired specimens with both pPNA types; these form clusters, except paired-specimens (E-143-7, E143-7), (E-142-5, E142-5), and (E-71-3, E71-3) that are in positive and negative axes. Axis 1 explains the microbial variability in all paired specimens and specimens from {\it Sonchus oleraceus } and {\it Sonchus arvensis} plants with highly abundant {\it Erysipelotrichia}. We scale the axis of ordination plots according to the eigenvalues to represent relative distances between specimens as faithfully as possible. It seems that the endosphere microbial variability comes from differences in plant types and specimens, much less from the amplification methods. Supplementary Figure \ref{fig1Vis} (B) suggests that specimens amplified with both types of pPNAs are composed of ASVs from different Classes. To enhance the visualizations of microbial count data, we can incorporate some of the phylogenetic information into the ordination summaries as described in the next section.

\subsection{Integrating the phylogenetic tree into the analyses}\label{phylo}

When considering the abundance of the different bacteria associated with the denoised ASVs identified through pipelines such as DADA2 \citep{DADA2}, strain variants are identified using standardized bacterial taxonomic databases \citep{RDP,silva} and phylogenies such as {\tt greengenes} \citep{Greengenes}. 

These phylogenetic relationships create a family tree, of which the tips
are the different taxa or strains (ASVs are also called Operational Taxonomic Units, OTUs). Thus, the count table row identifiers in Table \ref{tab01} are not evolutionarily independent, and there can be some benefit to taking this into account. Using the phylogenetic tree to inform distances or kernels between the abundance vectors was developed in \cite{purdom2011analysis}, extending the idea of doing a double principal coordinate analyses for ecological data as presented in \cite{Pavoine:2004}. 

\begin{figure}[H]
	\includegraphics[width=\linewidth]{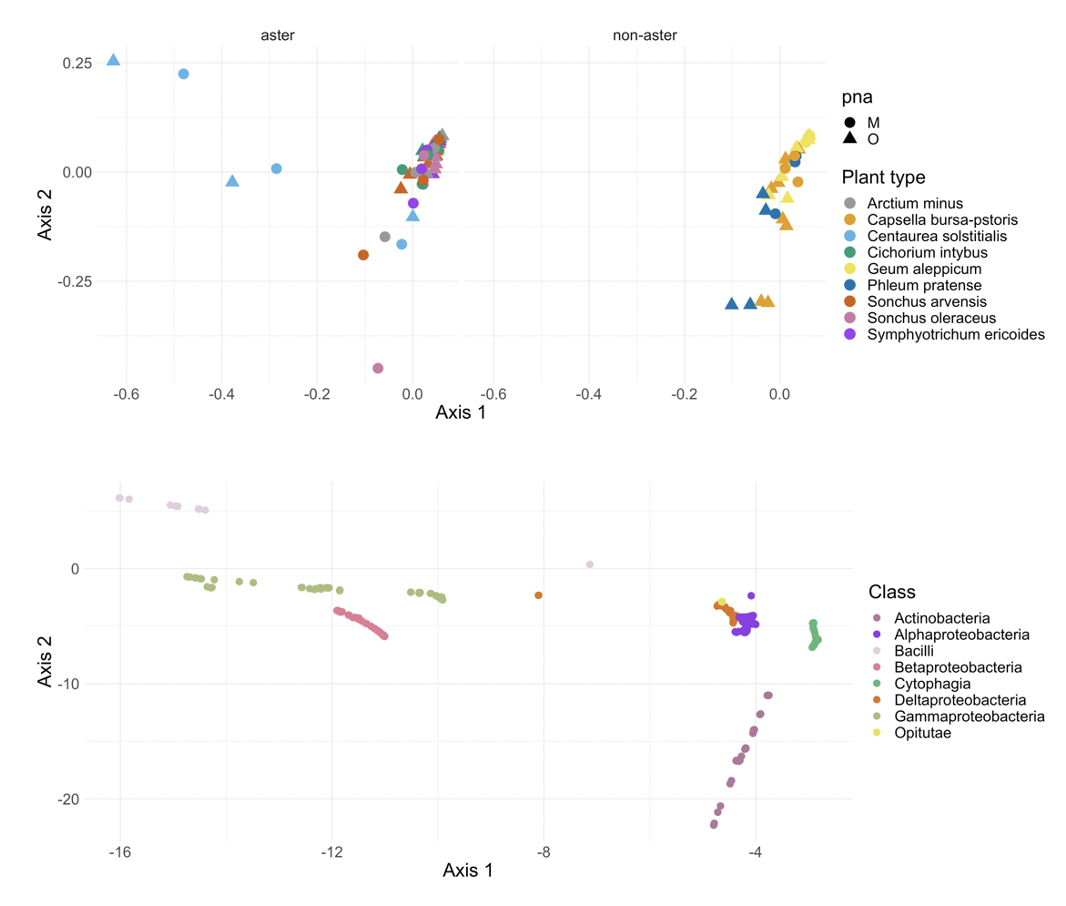}
	\caption{The results from adaptive gPCA reveal {\it Centaurea solstitialis} specimens are outliers among Asteraceae plants in the left of Axis 1. Axis 2 explains the microbial variability in plant types.}
	\label{agPCA}
\end{figure}

One difficulty in tree-based analyses is that the phylogenetic relationships between taxa may only explain a small percentage of the abundance differences between specimens (or beta-diversity, as it is called). More recent work has shown that a modulated penalized-tree approach allows a more nuanced use of the phylogenetic tree to interpret sample differences \citep{fukuyama2019adaptive,Fukuyama2017-c}. Testing procedures can help delineate what the tree can explain and how the residuals from the tree-based variability relate to the other specimen covariates. Figure \ref{agPCA} shows the results from adaptive generalized PCA (adaptive gPCA), available in {\tt adaptiveGPCA} package. It reveals {\it Centaurea solstitialis} specimens are outliers among Asteraceae plants on the left of Axis 1. Axis 2 explains the microbial variability in plant types. Like truncated-rank PCA, the paired-specimens in Supplementary Table \ref{tab04_paired} are relatively close to each other.

When testing for differential abundances between taxa, the unifrac score \textemdash a phylogenetic-based distance \citep{lozupone2005unifrac} or modifications thereof enable tree-based tests (for a review and comparisons between different tree-based distances see \cite{Fukuyama2012}). There have been several follow-up methods that use the development of phylogeny-based kernels to understand microbial diversity \citep{zhao2015testing,washburne2019phylofactorization}. There are still large areas that require additional research when it comes to propagating the uncertainties with which the phylogenies are known and the uncertainties with which the taxa are identified or the number of reads measured.

Several attempts have been made to leverage hierarchical Dirichlet processes to link the evolutionary processes at work with the ecological context. In particular, \cite{harris2015linking} consider the result of Hubbell showing that under the neutrality assumption, the abundances within the neutral guild fluctuates and that the number of species at a single site is a balance between the immigration of new species and local extinctions. 

\subsection{Correspondence analysis}
Correspondence analysis (CA) is a weighted bilinear method that provides a representation of a contingency table in a low-dimensional space \citep{Greenacre2010CA}. CA can be understood as a generalized singular value decomposition that provides factor scores for columns and rows of the contingency table that are then used to represent the association between rows (taxa) and columns (specimens) (Holmes, 2008). \nocite{Holmes2008}

\begin{figure}[H]
	\centering
	\includegraphics[width=\linewidth]{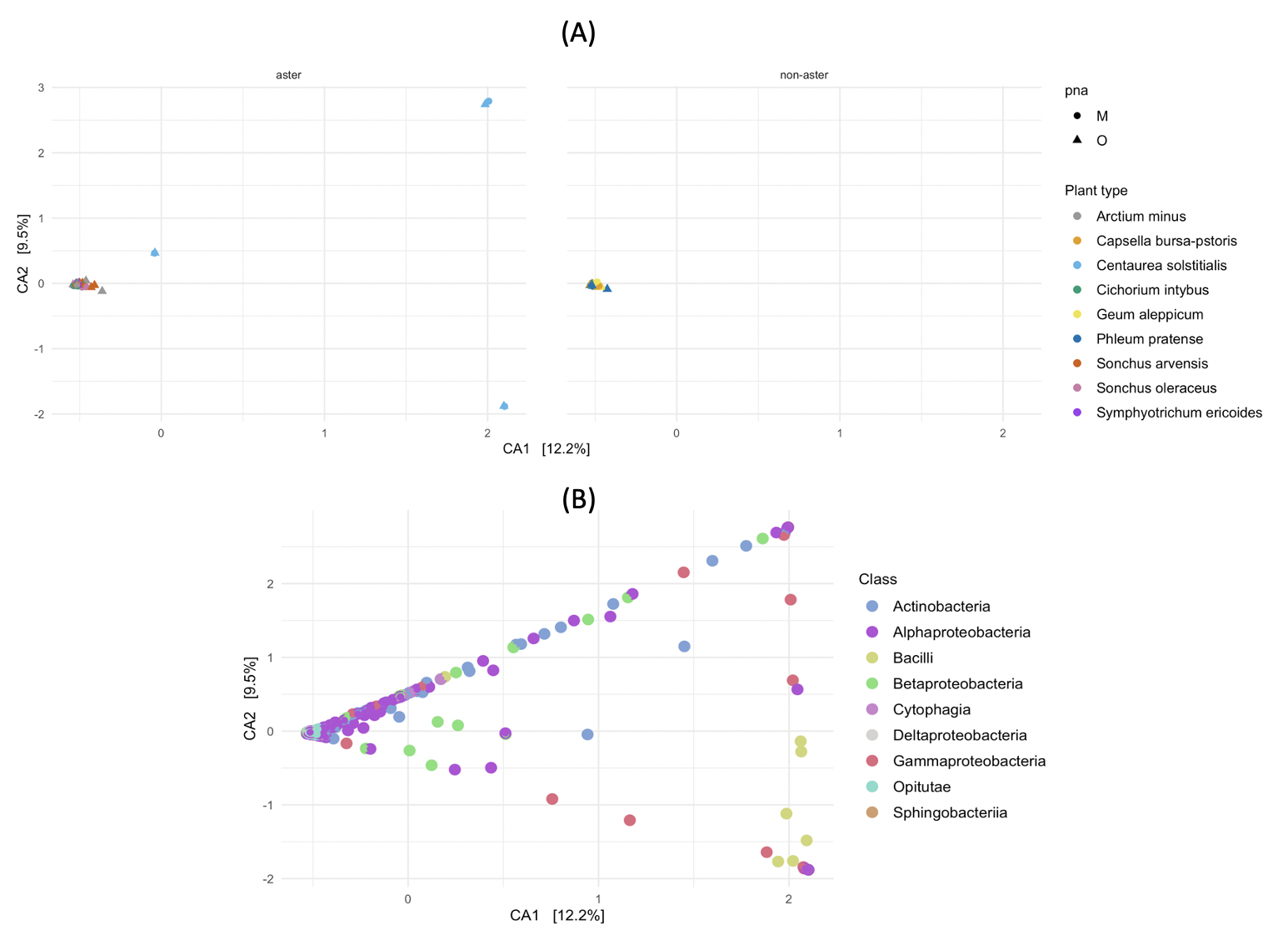}
	\caption{Correspondence analysis of all plant specimens. (A) plot shows the points of specimens and (B) plot shows the points of ASVs and the corresponding Class labels.}
	\label{fig01CA}
\end{figure}

CA uses chi-square distances, which are sensitive to outliers. The scree plot shows that the dimensionality of the underlying variation is two dimensional, and we see the first two dimensions explain about 21.7\% of the inertia (proportional to the sum of the chi-square distances). Figure \ref{fig01CA} shows that CA is not robust to the outlier {\it Centaurea solstitialis}. We observe five clusters of specimens apparent in the data. {\it Bacilli} contributes more to the cluster of one paired {\it Centaurea solstitialis} in the first axis, {\it Sphingobacteriia}, {\it Gammaproteobacteria}, {\it Betaproteobacteria}, and {\it Actinobacteria} dominate in a cluster of another paired {\it Centaurea solstitialis}, and other ASVs comprise the other three clusters of all other specimens.

We see that several combinations of distances and multivariate methods provide similar conclusions at ``larger levels."

\subsection{Network analysis}

We subset the 18 paired-specimens to perform a graph-based test. Supplementary Figure \ref{fig1net} shows the network based on fixing a maximum threshold for the Jaccard dissimilarity matrix. All paired-specimens are connected, except (E-143-7, E143-7), (E-142-5, E142-5), (E-71-2, E71-2), and (E-71-3, E71-3). {\it Centaurea solstitialis} paired-specimens, and all other paired-specimens make subgraphs. We performed a test using the minimum spanning tree (MST) built with the Jaccard dissimilarity (without thresholding). The null hypothesis is that the two pPNA types (universal and Asteraceae-modified) have the same microbial distribution. This test has a larger p-value (0.872); thus, we do not reject the null hypothesis. This result mirrors the observations from the PCA after the truncated-rank transformation, adaptive gPCA, MDS, and DPCoA that paired-specimens in Supplementary Table \ref{tab04_paired} have similar microbial variability.

\section{Differential Abundance Analysis}\label{diff}

An important goal in microbiome research is often to find taxonomic differences across environments or groups. Although strain switching may obscure the interpretation of these differences, if the strains stay the same, an adaptation of differential abundance techniques used in RNA-seq has proved useful. After visualizing the data, several possible downstream analyses are commonly used depending on the experimental design and questions of interest. For instance, consider a study that presents several arms, such as treatments and control. The most straightforward approach is to rank taxa according to how differentially abundant or differentially prevalent they are in the two conditions. Then, we follow a standard decomposition of variability according to random effects or fixed effects. In human studies, the random effects could correspond to subjects and  fixed effects to treatments.

Differential abundance analysis can be done on count data \citep{mcmurdie2014, love2014, robinson2010edger} or transformed data \citep{smyth2005limma}. In the latter case, it can be necessary to add a pseudo count for zeros.  It is preferable to do goodness of fit test to model counts for each taxon directly. 

\subsection{Permutation tests using distance matrices}
\label{permanova}
%The functions partition sums of squares of a multivariate data set, and they are directly analogous to MANOVA (multivariate analysis of variance). McArdle and Anderson (2001) and Anderson (2001) refer to the method as “permutational manova” (formerly “nonparametric manova”). Further, as the inputs are linear predictors, and a response matrix of an arbitrary number of columns, they are a robust alternative to both parametric MANOVA and to ordination methods for describing how variation is attributed to different experimental treatments or uncontrolled covariates. Functions are also analogous to distance-based redundancy analysis in functions dbrda and capscale (Legendre and Anderson 1999). Functions provide an alternative to AMOVA (nested analysis of molecular variance, Excoffier, Smouse, and Quattro, 1992; amova in the ade4 package) for both crossed and nested factors.

A popular confirmatory analysis  starts with the computation of dissimilarities between the samples using either the Jaccard, Bray Curtis, unweighted or weighted 
unifrac \footnote{the unifrac distance is a modification of the Wasserstein  distance computed along the phylogenetic tree\cite{Fukuyama2012,lozupone2005unifrac,Evans2012phylogenetic}.} distances. Then, under the null assumption that the sample abundance vectors 
are independent and identically distributed across groups, a null permutation distribution of a statistic dependent on the distances
can be compared to the  statistic calculated on the observed data.
Examples of such procedures include the permutational multivariate analysis of variance
{\bf \sc "permanova"}, introduced in \cite{anderson2001new} and available  as the {\tt adonis} function in the {\tt vegan} package for instance \cite{vegan2020}.
However the method is dependent on the assumption that the samples are (conditionally) independent and permutations of the full set give false positives if
the samples have nested a priori factors, in which case modified permutation procedures such as those suggested
in \cite{Excoffier1992analysis} are more appropriate. 

Some asymptotic theory is available for {\sc permanova} in the most restrictive cases of weighted Euclidean distances between independent samples \cite{anderson2003generalized}.
These tests
are sensitive to latent groupings and correlations between ASVs leading to common occurrences of false positives
and over-interpretation of the significance of the differences between groups. 
Distance based tests agglomerate the different ASVs into one dissimilarity index so they do not provide
indications as to which ASVs differ. One difficulty with current 16S rRNA data are that the 
distance based tests are very sensitive to the {\em choice} of distance and the presence of {\em strain switching} can substantially decrease the power of the test.
As an example, in 
the article --- from which we drew the data~\cite{fitzpatrick2018}---  the authors use  {\sc permanova} and report that they do not detect a difference between specimens grouped according to pPNA types (universal and Asteraceae-modified pPNA types). In this case, one explanation is that these types of 
permutation tests  are particularly underpowered  when one set of samples have one strain (ASV) and another
a slightly different strain, registered as a different ASV. In the supplementary material, we show that for the exemplary data this is in
fact the case as strains ASV 153 is switched with  ASVs 12, 354
and 345. 
To illustrate the decrease in power through a simulation, we
generated negative binomial count data with parameters similar to these ASVs
and show that when a species switches ASV, ie is present as one ASV in one set of samples and a close, distinct strain appears in
the other set of samples, the power to detect a difference using a Bray Curtis distance and {\sc permanova} is  considerably diminished
(see the code at \url{https://pratheepaj.github.io/diffTop/articles/appendix/08_differential_abundance_analysis.html} for the illustrative power calculations).

\subsection{Differential abundance through generalized linear modeling and transformations}

The bacterial abundances vary between subjects and environments
both because the underlying prevalences are different and because the sampling depths vary. The sampling depth variation causes
the count data to have unequal variances (heteroscedasticity). Mixture/hierarchical models are useful for this type of data.  Using a Poisson-gamma hierarchy achieves a first
model that results in negative binomial
\cite{mcmurdie2014} count data  for which the  generalized linear models for differential abundance analysis of microbiome data \citep{love2014, robinson2010edger} is well adapted. After an appropriate transformation on the count as a response, these generalized linear models can to detect the differences in bacterial abundances \citep{smyth2005limma}.
Testing can be done at different levels in the phylogenetic tree and adjustment for nested testing is available through packages such as   {\tt structSSI} that implements a multiple testing procedure that accounts for hierarchical dependence in the taxonomy table or phylogenetic tree \citep{Sankaran2014-vz}.

Environmental differences, weather change, animal or human interaction can create substantial heterogeneity in bacterial communities within and between subjects or locations. Therefore, longitudinal experiments are often preferred because they account for the within and between-subject/location variability. We have developed a moving block bootstrap method for differential abundance analysis in longitudinal studies \citep{jeganathan2018block} that accounts for the added dependences. This method resamples overlapping blocks of specimens within each subject to approximate the test statistic's distribution and tailors the pivoting procedure and block sizes to the data. 

Some microbiome studies incorporate spatial dependence into the differential abundance analysis along spatial gradients \citep{proctor2017landscape, proctor2018spatial}. \cite{singh2019nonparametric} proposed a nonparametric test to identify the effect of environmental factors that shape microbiome variability. This test can account for spatial dependence and inter-dependence among taxa.

For designed experiments, \cite{grantham2020mimix} developed a Bayesian mixed-effects model for testing the effect of environmental and treatment factors on microbiome variability. This method models the taxonomic counts as a multinomial and accounts for the interdependence between taxa by applying a hierarchical mixed-effects model.

Unfortunately, simple two-sample testing is marred by several technological difficulties. There are batch effects, technical biases, and heteroscedasticity in the prevalence estimates due to differences in the library sizes across specimens as well as strain switching (where ASV strains are replaced as we change locations or subjects). The taxonomic strains are often not pre-specified, requiring a nonparametric infinite-dimensional model and precluding the use of methods that assume a small well-defined set of categories ---  the case of compositional data methods---  for instance. In general, we recommend hierarchical models that can account for undetected taxa and heterogeneity of microbial distribution in specimens.

\subsection{Communities instead of individual taxa}
%As we saw above, strain switches poses problems when identifying differentially abundant taxa.
As noted above, high-resolution acquisition of data at the taxonomic strain level resulted in different subjects or environments presenting slightly different taxa that are ``functionally synonymous"; thus we have to deal with these {\em strain switches}. The individual taxa may not be as important as their combination. This makes the problem similar to how synonyms occur in textual analyses. The co-occurrence of bacteria and the departures from a simple multinomial model make the analogy between textual and microbiome data analysis useful. \cite{Sankaran2018-LDA} demonstrated the utility of the analogy between textual analysis and molecular microbial ecology.
In textual analyses, documents are of unequal lengths, and topics enable the simplification of document contents. Documents can have several topics; thus, mixed membership is the relevant model. The same is true in microbial ecology, where several communities can be present in a specimen. A probability distribution over bacteria characterizes each community. This has the advantage over the simpler Dirichlet multinomial mixture models \cite{holmes2012dirichlet}
%(Holmes et. al., 2012) 
that bacteria can be very interdependent. Syntrophy occurs when one bacterium cannot survive without another and can be modeled by having a simple two bacteria topic, whereas syntrophy is impossible to model with a simple multinomial. In the next section, we describe topic models and illustrate their use on our example data set. In particular, topic models provide useful aggregates that can be used for differential abundance analysis based on topics rather than individual strains. \nocite{holmes2012dirichlet}

\section{Latent Dirichlet Allocation} \label{LDA}

We use the latent Dirichlet allocation (LDA) model as described by \cite{Sankaran2018-LDA}. We suppose that $\vK_{.j}$ denotes the $m \times 1$ vector of taxa abundance in specimen $j,$ where $j =1, 2, \cdots, n_{1}$ and $T$ is a prespecified number of topics. The LDA generative process for each specimen $\vK_{.j}$ of size $S_j$ follows the following steps: Each specimen is associated to a set of  topics drawn from a probability distribution $\vtheta_{j} \overset{\text{iid}}{\sim} \text{Dirichlet}_{T}\left(\valpha \right)$, over a mixture of latent topics (bacterial communities). Each topic is associated with a probability distribution $\vbeta_{t} \overset{\text{iid}}{\sim} \text{Dirichlet}_{m}\left(\vgamma \right)$ over taxa. We draw $\vK_{.j}$ from $\left. \vK_{.j} \right \vert S_{j}, \vtheta_{j} , \mB\overset{\text{iid}}{\sim} \text{Multinomial}\left(S_{j},  \mB  \vtheta_{j} \right)$, where $j = 1, \cdots, n_{1}$, $\mB = \left[\vbeta_{1}, \cdots, \vbeta_{T}\right]^{T},$ and $S_{j}$ is the library size of specimen $j.$

We set the hyper-parameters $\valpha$ and $\vgamma$ less than one to generate sparse mixtures that are different from each other and avoid generating unrealistic topics. We estimated the model parameters using  Hamiltonian Monte Carlo and the No-U-Turn (HMC-NUTS) method implemented in Stan \citep{carpenter2017stan}. We denote the Bayesian posterior estimates $\lbrace \hat{\theta}_{j},  \hat{\vbeta}_{t}\rbrace, j= 1, 2, \cdots, n_{1}$ and $t = 1, 2, \cdots, T.$ 

For our example, we chose the number of topics $T=11$ based on an ordination analysis. For the data set in consideration, there were 1418 distinct ASVs across 86 specimens after selecting ASVs with at least 25 reads in at least two specimens. Ordination plots show that bacterial signatures vary between Asteraceae and non-Asteraceae plants, {\it Centaurea solstitialis} specimens are outliers among Asteraceae plants, and the paired specimens are relatively similar in microbial variation. For eight different plants and {\it Centaurea solstitialis} from three other countries, we chose the number of topics $T = 11$. We set $\alpha$ to be 0.8 across all 86 specimens and  $\gamma$ to be 0.5  across all ASVs. We estimated the parameters using the HMC-NUTS sampler with four chains and 2000 iterations. Out of these 2000 iterations, 1000 iterations were used as warmup samples and discarded. 
Label switching across chains makes it difficult to directly compute log predictive density, split-$\hat{R}$ \citep{vehtari2020rank, gelman1992inference}, effective sample size  \citep{kruschke2014doing} for model assessment, and evaluate convergence and mixing of chains. To address this issue, we fixed the order of topics in chain one and then found the permutation that best aligned the topics across all four chains. For each chain two to four, we identified the estimated topics pair with the highest correlation, then found the next highest pair among the remaining, and so forth. 

We present the predictive model check for $T=11$, with simulated and observed data using a statistic $G\left(K_{ij} \right) = \text{max}\lbrace K_{ij} \rbrace.$ Supplementary Figure \ref{fig_ppc_all_samples} shows the histograms of $G\left(K_{ij} \right)$ of each ASV in simulated data from the fitted model and the horizontal line of the observed $G\left(K_{ij} \right)$. According to the histograms, the LDA model with eleven topics makes a realistic prediction. Supplementary Figures \ref{fig_ess_all_samples} and \ref{fig_rhat_all_samples} show the effective sample size (ESS) and split $\hat{R}$ with eleven topics. These diagnostics provide some evidence of good mixing and convergence of the chains to the stationary distribution. 

The host contamination plastid is significantly reduced in Asteraceae plants with M-pPNA type \citep{fitzpatrick2018}. But only four plants {\it Sonchus arvensis}, {\it Arctium minus}, {\it Sonchus oleraceus}, {\it Centaurea solstitialis} of Asteraceae have paired-specimens. Now, we infer whether the pPNA types affect microbial distributions in different plants. In endosphere specimens, bacteria that have similar functionalities in each plant type can co-occur as latent communities, and topic modeling provides insight into microbial communities across these types. With the goodness-of-fit of the LDA model, we can draw an informative summary of bacterial communities in each plant type with O and  M-pPNA types. Hence after the topic analysis, we choose to study one plant type at a time. The analogous figures for the other plant types are available as Supplementary Figures \ref{fig_topic_centaurea_solsti_K_11} and \ref{fig_topic_dis_all_K_11}. 

\begin{figure}[H]
	\centering
	\includegraphics[width=\linewidth]{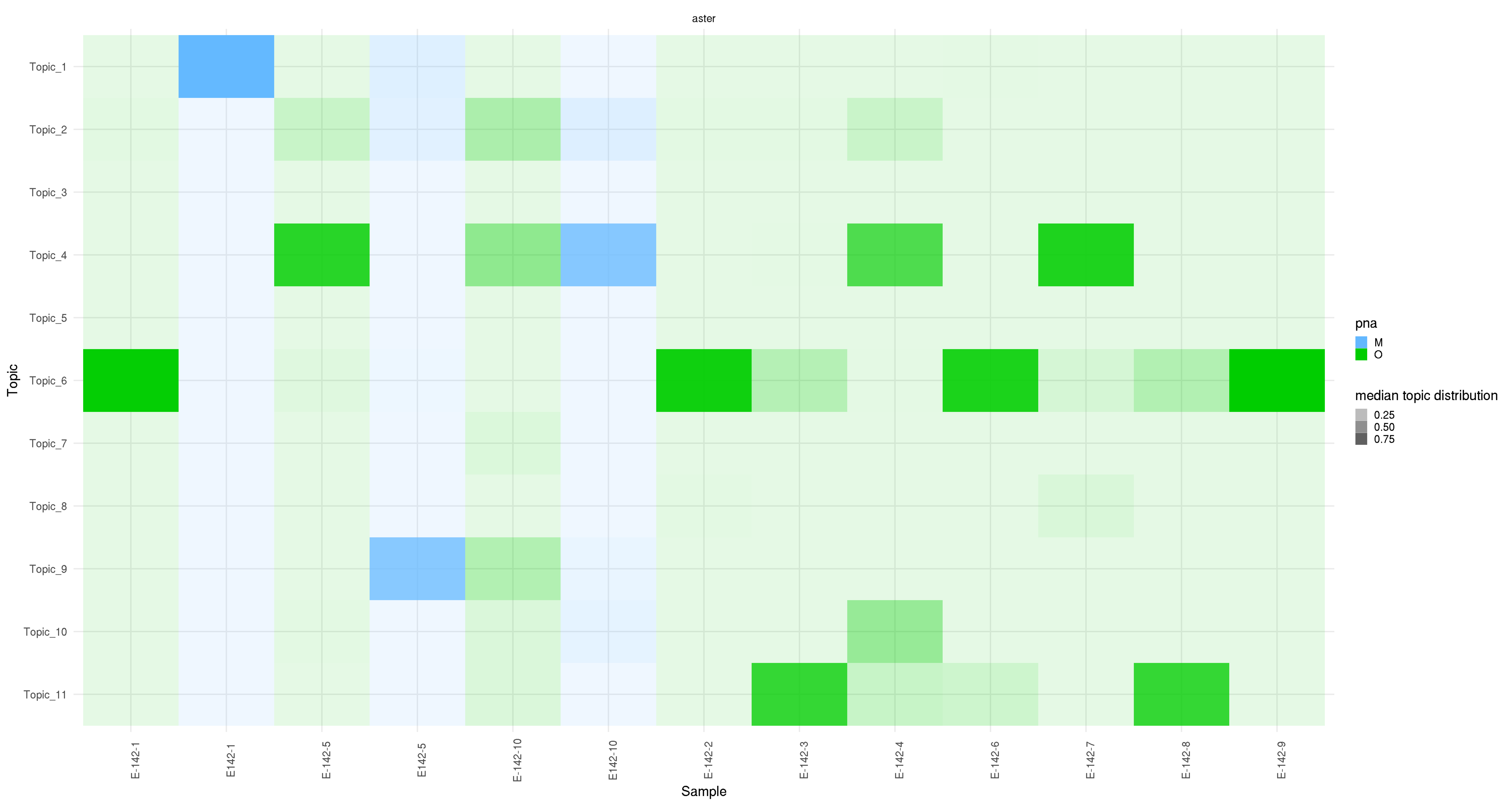}
	\caption{Topic distribution in specimens from {\it Sonchus Arvensis} plant. The color gradient represents the median topic distribution.}
	\label{fig_topic_dis_sonchus_arven_K_11}
\end{figure}

\begin{figure}[H]
	\centering
	\includegraphics[width=\linewidth]{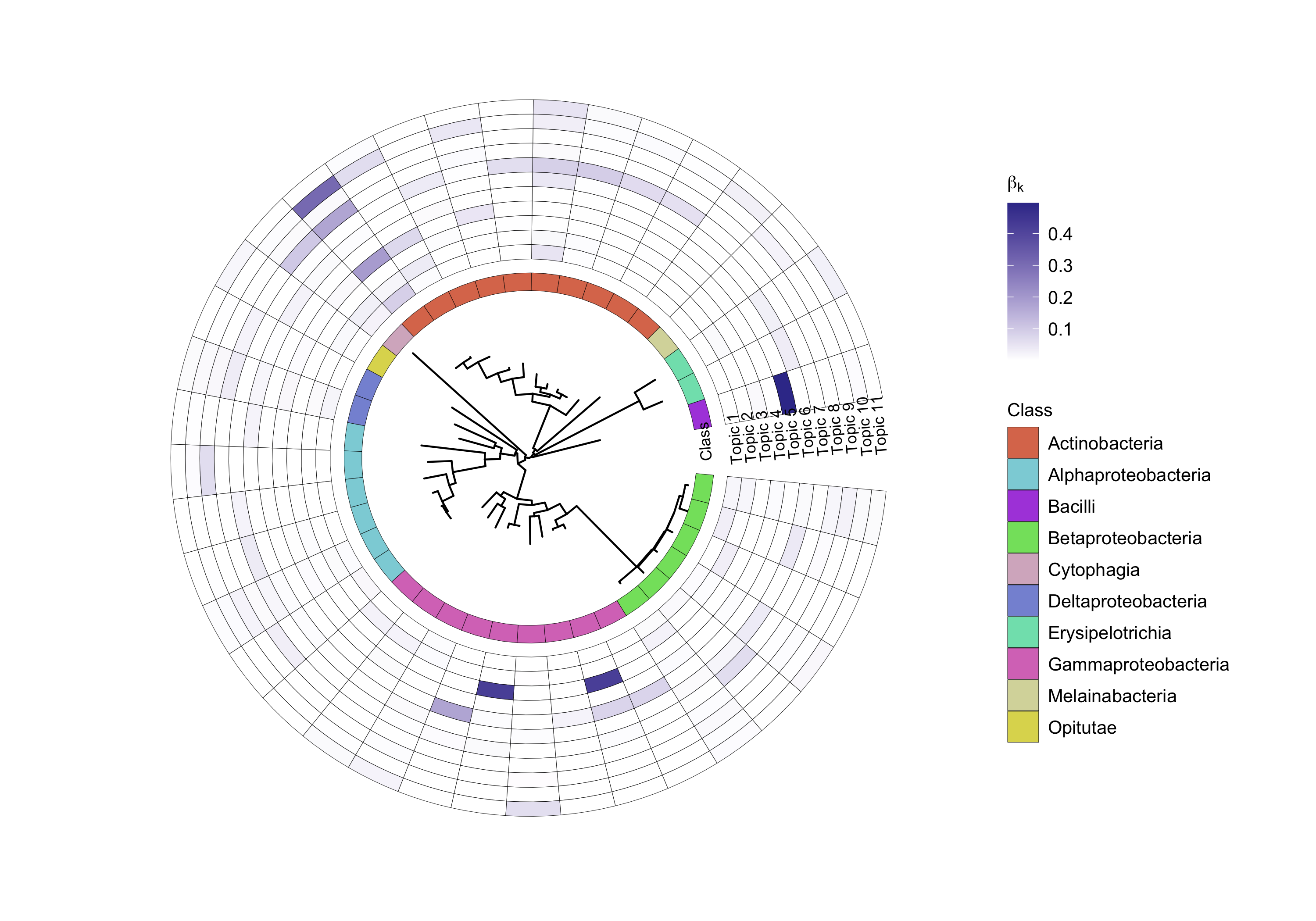}
	\caption{ASV distribution over topics in all specimens.}
	\label{fig08_asv_dis_K_11}
\end{figure}

Figure \ref{fig_topic_dis_sonchus_arven_K_11} shows the topic distribution in each specimen from the {\it Sonchus Arvensis} plant. Among the eleven topics for all plants, seven topics were predominately present in specimens from  {\it Sonchus Arvensis}, an Asteraceae plant with 13 replicates specimens. Ten specimens were amplified with O and three with M-pPNA types, respectively. Among these specimens, E-142-1, E-142-5, and E-142-10 are paired because both pPNAs were used for the same DNA specimens. Paired-specimens (E-142-1, E142-1) and (E-142-5, E142-5) have a different mixture of topics. Figure \ref{fig08_asv_dis_K_11} shows the ASVs distribution for each topic. Paired-specimen E-142-1 has largest proportion of Topic 6, which has distribution over ASVs from Classes {\it Acitinobacteria}, {\it Betaproteobacteria}, and \textit{Eryscpelotrichia} whereas other paired-specimen E142-1 has largest proportion of Topic 1, which has similar ASV distribution, except of Class \textit{Eryscpelotrichia}. Similarly, some paired-specimens from {\it Arctium minus} and {\it Sonchus oleraceus} plants have different mixtures of topics (Supplementary Figure \ref{fig_topic_dis_all_K_11}). The replicates of \textit{Centaurea solstitiali} that sampled across three countries show variation in microbial distribution, but the paired-specimens have similar distributions (Supplementary Figure \ref{fig_topic_centaurea_solsti_K_11}).

We use the microbial community-topics in each specimen to do a differential topic analysis to test whether topic memberships differ across conditions.
First, we compute the median Bayesian posterior of topic proportions in each specimen. Then, we multiply the proportions by the library size and round to an integer. This will give an abundance of the topic in each specimen. The we applied {\tt DESeq2} to identify differentially abundant topics across O and M pPNA types. Supplementary Table \ref{tab_diffT_all} shows that Topics 4, 5 and 10 are differentially abundant in O and M pPNA types. We also consider testing on only paired-specimens from Asteraceae and non-Asteraceae plants. Supplementary Figure \ref{fig_topic_dis_all_K_11} shows the large proportion of Topics 1, 9,  and 11 in Asteraceae paired-specimens. Supplementary Table \ref{tab_diffT_all_aster_paired} shows that these microbial communities are different in pPNA types. Topics 1, 2, 8, 9, 10 dominates in non-Asteraceae paired-specimens, and Supplementary Table \ref{tab_diffT_all_nonaster_paired} shows that we do not reject the hypothesis that these topics are differentially abundant. We conclude that microbial communities are significantly different in some Asteraceae plants. 

Finally, we note that the choice of the number of topics is possible through a Bayesian nonparametric approach that incorporates an infinite number of topics as described by \cite{blei2010probabilistic}.  Sources of biological variability can be incorporated into the hierarchical model; subject or location variation can be modeled as random effects. Other covariates, experimental arm variability (for instance, one group of subjects/locations may be treated) can be easily added. In the study of the human microbiome, it is common to include a subject's age or the level of urbanization in the food supply at a given location \citep{yatsunenko2012human}. 

An extension to the topic models described above  is provided by  a Bayesian nonparametric factor models that accomodates changing distributions of unbounded numbers of taxa in specimens. These models decompose the biological variation in bacterial communities' composition in a low-dimensional space defined along {\em continuous} latent factors rather than through {\em discrete} topics \citep{ren2017bayesian}. This method enables the propagation of uncertainty from the microbiome data to their multivariate ordination, and we illustrate this on our example data set below. 

\section{Uncertainty quantification for  ordination analysis}\label{BN}

\cite{ren2017bayesian}'s nonparametric Bayesian approach provides one way of modeling infinitely many possible taxa and specimen-specific microbial distributions. Their approach assumes an underlying finite-dimensional factor model that represents dependencies and uses a kernel decomposition  of the underlying communities. If we let unknown specimen-specific microbial distributions be $P^{j}\lbrace Z_{i} \rbrace, j =  1, \cdots, n_{1}$ and $Z _{i}$ be the i-th taxon of unbounded set, then Gram matrix $\phi\left(j_{1}, j_{2}\right)_{j_{1}, j_{2} \in \lbrace 1, \cdots, n_{1}\rbrace}$ defines the similarity between microbial distributions in specimens $j_{1}$ and $j_{2}$. \cite{ren2017bayesian} represent in the prior the similarity between $P^{j}$ through latent factor model in low dimensional space $\vY^{j}$ and use the posterior samples of normalized Gram matrix for the ordination analyses. This Bayesian approach has the advantage of providing posterior probability credible sets.

\begin{figure}[H]
	\centering
	\includegraphics[width=\linewidth]{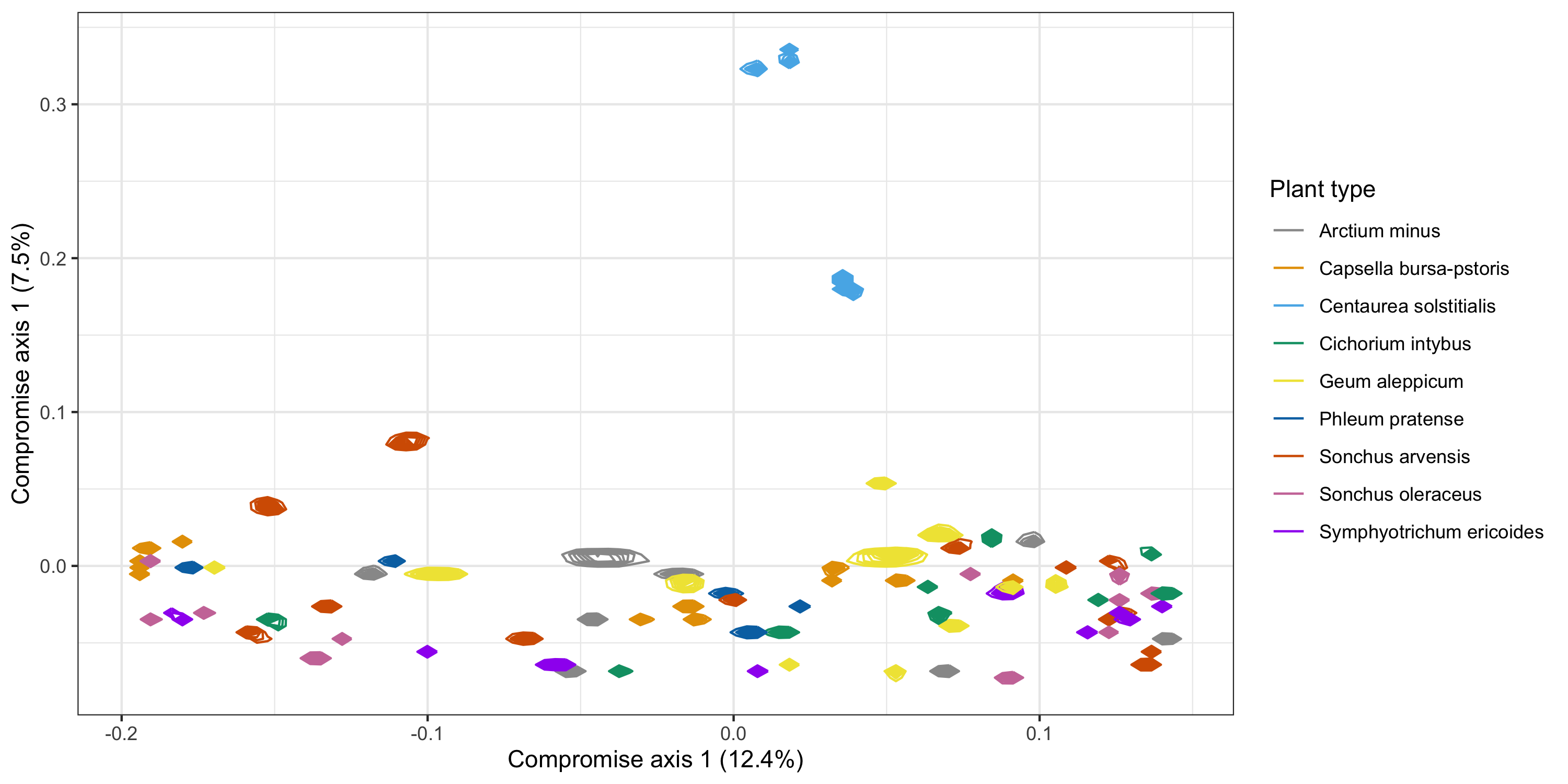}
	\caption{Ordination plot of specimens and 95\% posterior credible regions. We plot the first two consensus axes. The percentages on the two axes are the ratios of the corresponding eigenvalues and the trace of the matrix. Color indicate the plant type. }
	\label{BNO}
\end{figure}

We reused this technique on our example data set, where we choose ASVs that occur in at least eight specimens with at least 100 reads. This filtering resulted in 152 ASVs in 86 specimens. We start with 86 latent variables that assume one factor for each specimen. Then, we used Gibbs sampling with 50,000 iterations and a thinning size ten. Figure \ref{BNO} shows Bayesian nonparametric ordination for all the specimens. The contours represent the variability in the position of each specimen in the consensus space. Since we filtered the data down to very consistent ASVs, the credible regions for all specimens are relatively small. The two paired specimens from {\it Centaurea Solstitialis} are outliers that corroborate the exploratory analysis results with credible regions. 

\section{Discussion}
We have reused the 16S rRNA sequencing data from \cite{fitzpatrick2018} to review and demonstrate goodness of fit tests, multiplicative and additive models, variance-stabilizing and truncated-rank transformations, multivariate methods, network analysis, differential topic analysis, and uncertainty quantification for ordination analysis of sequencing reads in microbial ecology. The study used the 16S rRNA gene sequencing with universal (O) and Asteraceae-modified (M) plastid peptide nucleic acids (pPNA) types that limit the amplification of host-derived DNA, such as plastids, which share existing similar bacterial lineages. \cite{fitzpatrick2018} designed the experiment to test the efficacy of pPNA types in limiting the plastid contamination and seeing if there was a difference in microbial variation between pPNA types. By applying a linear mixed-effects model for each taxon, \cite{fitzpatrick2018} concluded that the pPNA types do not affect the identification of individual bacterial taxa and within specimen $\alpha-$  diversity for different metrics. The authors found little evidence of a difference between specimen $\beta-$diversity using permutational multivariate analysis of variance ({\sc permanova}) \citep{anderson2005permutational}. The original paper used a paired-designed experiment for some replicates of each plant. In this case, it would have been preferable to tailor the standard permutation method to the experimental design and incorporate ASVs that seemed to have switched labels. In general, strain switching precludes simple differential abundance analysis at the taxonomic level and we have introduced a more meaningful differential topic analysis that identifies  some topics in the Asteraceae paired-specimens that are significantly different. 

Bar plots and simple multinomial models have limitations in providing a complete picture of microbial variability, especially when syntropic relations create dependence between bacterial occurrences. The phylogenetic relationships between taxa can be important for biological interpretation, and interactive tools can enhance our understanding of the distribution of different taxonomies \citep{mcmurdie2015shiny,sankaran2018interactive}. If we use sorted prevalence to visualize each taxonomy, we may  lose specific rare bacteria present in the data, so it is important to do both analyses with presence absence data and that of the core microbiome with abundances. Heatmaps can identify clusters of specimens and the contribution of ASVs for each cluster, although a large number of ASVs  can make this difficult to do in one step.

Our review shows that ordination based on different distances can incorporate phylogenetic information, or incorporate sample depth weights as in correspondence analysis. For the example at hand, multidimensional scaling (MDS, PCoA) and correspondence analysis (CA) show that only a few paired-specimens from the Asteraceae plants and non-Asteraceae  form clusters. 

Among dimensional reduction methods, tree based metric ordination such as double principal coordinate analysis (DPCoA) is more interpretable and here it uncovers the dominant taxa in each cluster. These ordination plots depict only the most abundant or prevalent ASVs in clusters of specimens. As an alternative,
topic models are shown to be the more interpretable as they unmask rare and synonymous taxa and can enhance our understanding of 
the assembly of microbial communities  as topics which can be projected onto the phylogenetic tree. 
The ASV distributions over interpretable topics
demonstrate that differential topic analysis can enhance our understanding of the differences in complex microbiome communities (here across different plants and pPNA types).

In our reanalysis of the data, there were six Asteraceae plants and three non-Asteraceae plants with 18 paired-specimens, as in Section \ref{exa}. Compared to O-pPNA, the M-pPNA significantly reduced the host-derived DNA contamination in only Asteraceae plants. The topics we uncovered in the latent Dirichlet allocation (LDA) explained the differences in the mixture of topics in some paired-specimens of Asteraceae plants. Nevertheless, there were not enough paired-specimens to test the difference in microbial distribution. In addition, we found that eleven topics were distributed over mixture of ASVs from Classes {\it Acitinobacteria}, {\it Alphaproteobacteria}, {\it Betaproteobacteria}, {\it Deltaproteobacteria}, , {\it Gammaproteobacteria}, {\it Bacilli}, {\it Cytophagia}, {\it Erysipelotricha}, and {\it Opituate}. {\it Acitinobacteria} was not in Topics 3, 5, and 9, and two paired-replicates of {\it Centaurea solstitialis} has the largest proportion of Topics 3 and 5, which would be the reason for identifying {\it Centaurea solstitialis} specimens as outliers in ordination analyses. In contrast to results on Asteraceae plants, \cite{fitzpatrick2018} concluded that M-pPNA type did not limit deriving host-contamination in non-Asteraceae plants. We found fewer of topics in non-Asteraceae than Asteraceae plants as in Supplementary Figure \ref{fig_topic_dis_all_K_11}. Supplementary Table \ref{tab_diffT_all_nonaster_paired} shows that there is no evidence to have different mixtures of topics in O and M-pPNA types in non-Asteraceae plants. 

Latent microbial communities (topics) can share ASVs that are correlated. We could have enhanced our analyses by using
correlated topic models  \citep{blei2006correlated} that can be used if some topics are expected to be exclusive or negatively correlated. 
LDA does not provide the number of topics or account for covariates; for this, we recommend using hierarchical Dirichlet process topic modeling with a stick-breaking process for the base measure that can account for covariates and enable the inference of the number of topics from the data. 

Finally, we show how credible regions for ordination can help in the confirmatory phase of evaluating the relative similarity of specimens. Here, the Bayesian nonparametric ordination corroborated the LDA results on specimens showing dominant topics in most replicate specimens.

\section{Open problems and statistical challenges}

Microbiome data from designed experiments pose numerous statistical challenges because they are high-dimensional, heteroscedastic, and sparse. These count data are not normally distributed, and linear models that assume normality are not appropriate. Some transformations such as those in {\tt Voom-limma} \citep{smyth2005limma} have been proposed; these lead to the use of raw counts to count per million (CPM). Multiplying by large factors is tempting; however, the risk of invalidating the downstream statistical inferences is high. For instance, in the linear discriminant analysis effect size (LEfSe) method \citep{segata2011metagenomic} multiply relative abundances by a million; this creates artificially large sample sizes and bloats the power of the experiment resulting in a large false-positive rate.

Our exemplary data were cross-sectional with an ideal paired-sample design. Complete randomized design, randomized complete block design, split-plot design, longitudinal experimental designs, spatial and factorial designed experiments require much more intricate hierarchical mixture models. Because of the strong between location/subject variability, longitudinal designs are currently the most used in microbiome studies. The transformation and visualization methods that we discussed in this paper can be used for any exploratory analysis. For the differential abundance analyses, hierarchical generalized linear mixed models or moving block bootstrap~\citep{jeganathan2018block} have proved successful in the cases where the strains stay consistent across samples, but many more extensions need to be developed to incorporate more complex structured designs.

The tools we have shown here for analyzing marker-gene counts use standard statistical techniques such as nonparametric or parametric variance stabilization through Anscombe's transformation, generalized log transformation, and smoothing techniques. However, research is still needed on how to invert these transformations after the latent factors in the data have been discovered. Open questions include propagation and quantifying uncertainty in the phylogenetic relationships, assessment of the effect of technical biases, and count estimations on the downstream analytics, even in well defined generative models. 

Normalization and denoising methods for shotgun metagenomic tables are less developed than those for the marker-gene microbiome data; extra caution is necessary in accounting for the difference in gene sizes and the existence of pseudogenes and duplicate genes in the same genomes \citep{Quince2017shotgun}. Some recent efforts leverage probabilistic methods and use a few anchor genes, a similar procedure to the marker gene, see, for instance, \cite{Quince2017desman}, others use Bayesian hierarchical approaches to model subsets of short reads (called k-mers, see \cite{lu2017bracken}). Some methods do not explicitly align, assemble, or label reads, but simply embed the k-mers in a continuous space following modern unsupervised methods used in textual analysis \citep{menegaux2019continuous}, however inferential properties of these algorithms under realistic conditions are not understood. Recently, \cite{Quince2020metagenomics} devised a bioinformatics tool for high-resolution shotgun metagenomics (STRONG). It may be useful to apply the statistical tools reviewed in this paper to metagenomic data output from that pipeline.

The analytics we have illustrated here used R, Bioconductor packages, and Stan for reproducible microbiome data analyses \citep{team2013r, gentleman2004bioconductor, carpenter2017stan}. No such high-level statistical toolbox exists as yet for metagenomic data.

Evaluation of the robustness of inferences to the multiple choices of distances and filtering parameters has only been done empirically up to now. As an example, the choice in the number of topics used in the Latent Dirichlet allocation models could be done by using another level in the Bayesian hierarchical model, but appropriate prior distributions for topics for microbiome data depend strongly on exactly what type of environment is under study, and more calibration experiments are needed to guide the user in their choices. The choice of the prior on the number of topics is important in determining how bacterial diversity changes when the number of specimens increases, as is shown in our example dataset, where the paired-specimen design improved the results of the analyses. 

This review has only scratched the surface of the potential for the use of statistics in microbial ecology. \cite{mcmurdie2013} built the {\tt phyloseq} project as a bridge between bioinformatics pipelines and the statistical toolset enabling users to account for multivariate, spatiotemporal structure in the data. Opportunities abound for refining the basic analyses presented here: many problems have data with dependent sampling designs that benefit from the use of geostatistical methods, time series monitoring under perturbations, and ecological statistics with community assembly and network analyses. 

Bayesian approaches have practical advantages; future research directions would focus on Bayesian nonparametric models. To use these methods, it would be worthwhile for the community to publish enhanced cases studies using Bayesian inference in Stan \citep{carpenter2017stan} and using R/Bioconductor S4 structures and packages. Microbiome research relates tightly to spatiotemporal and single-cell 'omics data, and we hope this review enables statisticians to extend ideas for more sophisticated, reproducible, interpretable microbiome data analyses. 

Future work will certainly extend the existing statistical methods used in environmental and ecological studies to provide more complete experimental designs, hierarchical Bayesian models, and the incorporation of geostatistics and spatiotemporal structures. We see a bright future in applying Bayesian and hierarchical methods to analyze these data and look forward to seeing more statisticians involved in this area.

%%%%%%%%%%%%%%%%%%%%%%%%%%%%%%%%%%%%%%%%%%%%%%
%% Single Appendix:                         %%
%%%%%%%%%%%%%%%%%%%%%%%%%%%%%%%%%%%%%%%%%%%%%%
%\begin{appendix}
%\section*{???}%% if no title is needed, leave empty \section*{}.
%\end{appendix}
%%%%%%%%%%%%%%%%%%%%%%%%%%%%%%%%%%%%%%%%%%%%%%
%% Multiple Appendixes:                     %%
%%%%%%%%%%%%%%%%%%%%%%%%%%%%%%%%%%%%%%%%%%%%%%
%\begin{appendix}
%\section{???}
%
%\section{???}
%
%\end{appendix}

%%%%%%%%%%%%%%%%%%%%%%%%%%%%%%%%%%%%%%%%%%%%%%
%% Support information (funding), if any,   %%
%% should be provided in the                %%
%% Acknowledgements section.                %%
%%%%%%%%%%%%%%%%%%%%%%%%%%%%%%%%%%%%%%%%%%%%%%
 \section*{Acknowledgements}
% The authors would like to thank ...
% 
% The first author was supported by ...
% 
% The second author was supported in part by ...
We are grateful for the thoughtful reading and suggestions made by David Relman and his group, Brian Reich and the referees that helped improve the manuscript.
This work was funded by a VMRC grant from the Gates foundation and a grant R01AI112401 form the NIH. 
We are happy to acknowledge to the R and Bioconductor Core Teams and authors of the packages {\tt BARBI, dada2}, {\tt DESeq2, phyloseq, decontam, ggplot2, rstan, adaptiveGPCA}, {\tt tidyverse} which were used for constructing figures and running the analyses in this paper.

%%  imsart-nameyear.bst  will be used to                   %%
%%  MR numbers will be added by VTeX.                      %%
%%                                                         %%
%%  Use \cite{...} to cite references in text.             %%
%%                                                         %%
%%%%%%%%%%%%%%%%%%%%%%%%%%%%%%%%%%%%%%%%%%%%%%%%%%%%%%%%%%%%%
\newpage
%% if your bibliography is in bibtex format, uncomment commands:
\bibliographystyle{imsart-nameyear} % Style BST file
\bibliography{JABES_paper_arXiv}       % Bibliography file (usually '*.bib')

%% or include bibliography directly:
% \begin{thebibliography}{}
% \bibitem[\protect\citeauthoryear{???}{???}]{b1}
% \end{thebibliography}

%%%%%%%%%%%%%%%%%%%%%%%%%%%%%%%%%%%%%%%%%%%%%%
%% Supplementary Material, if any, should   %%
%% be provided in {supplement} environment  %%
%% with title and short description.        %%
%%%%%%%%%%%%%%%%%%%%%%%%%%%%%%%%%%%%%%%%%%%%%%
%\begin{supplement}
%\stitle{???}
%\sdescription{???.}
%\end{supplement}

\newpage
\begin{supplement}
\stitle{Figures and tables produced in the analyses.}
\sdescription{}
\end{supplement}

\setcounter{figure}{0} % starts numbering S F 1
\renewcommand{\figurename}{Supplementary Figure}
\setcounter{table}{0}
\renewcommand{\tablename}{Supplementary Table}

	\begin{figure}[H]
		\centering
		\includegraphics[width=\linewidth]{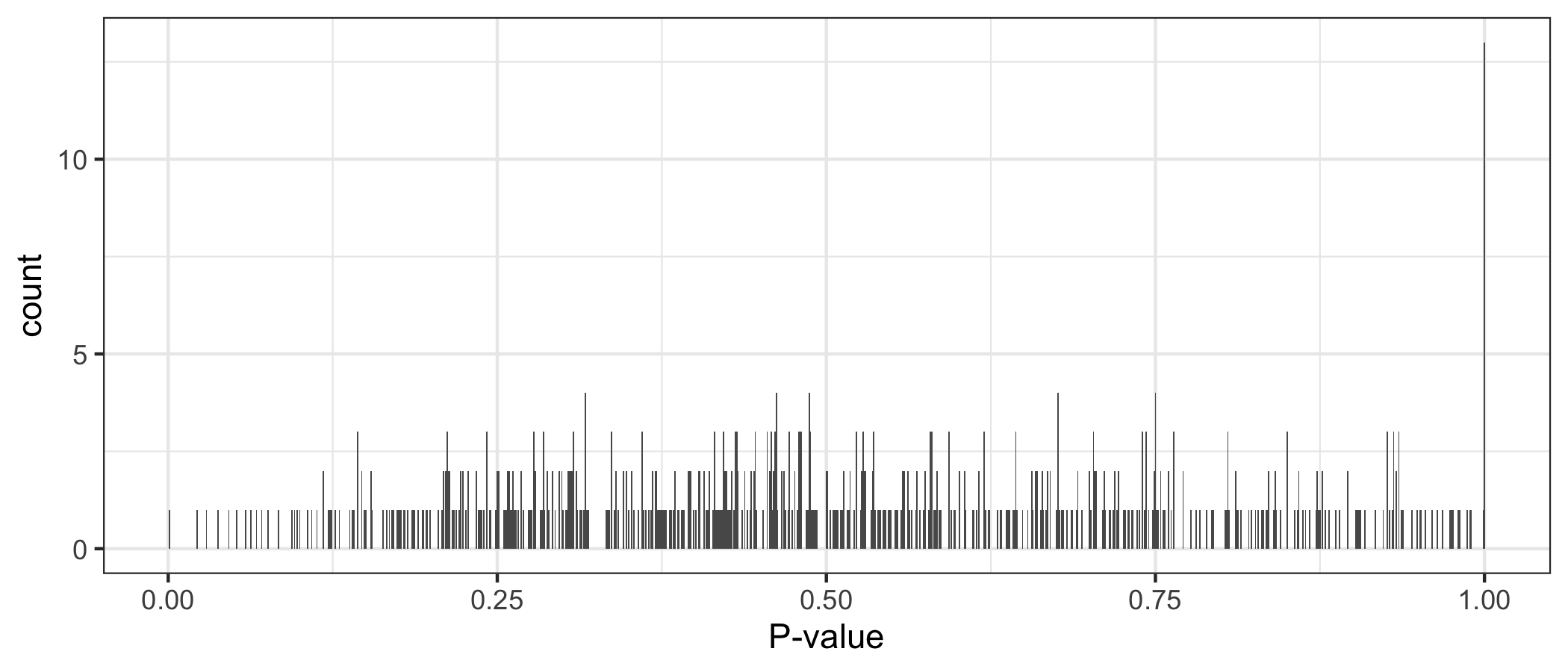}
		\caption{P values for goodness of fit test of negative binomial distribution for each taxon.}
		\label{fig02}
	\end{figure}

	\begin{figure}[H]
		\centering
		\includegraphics[width=\linewidth]{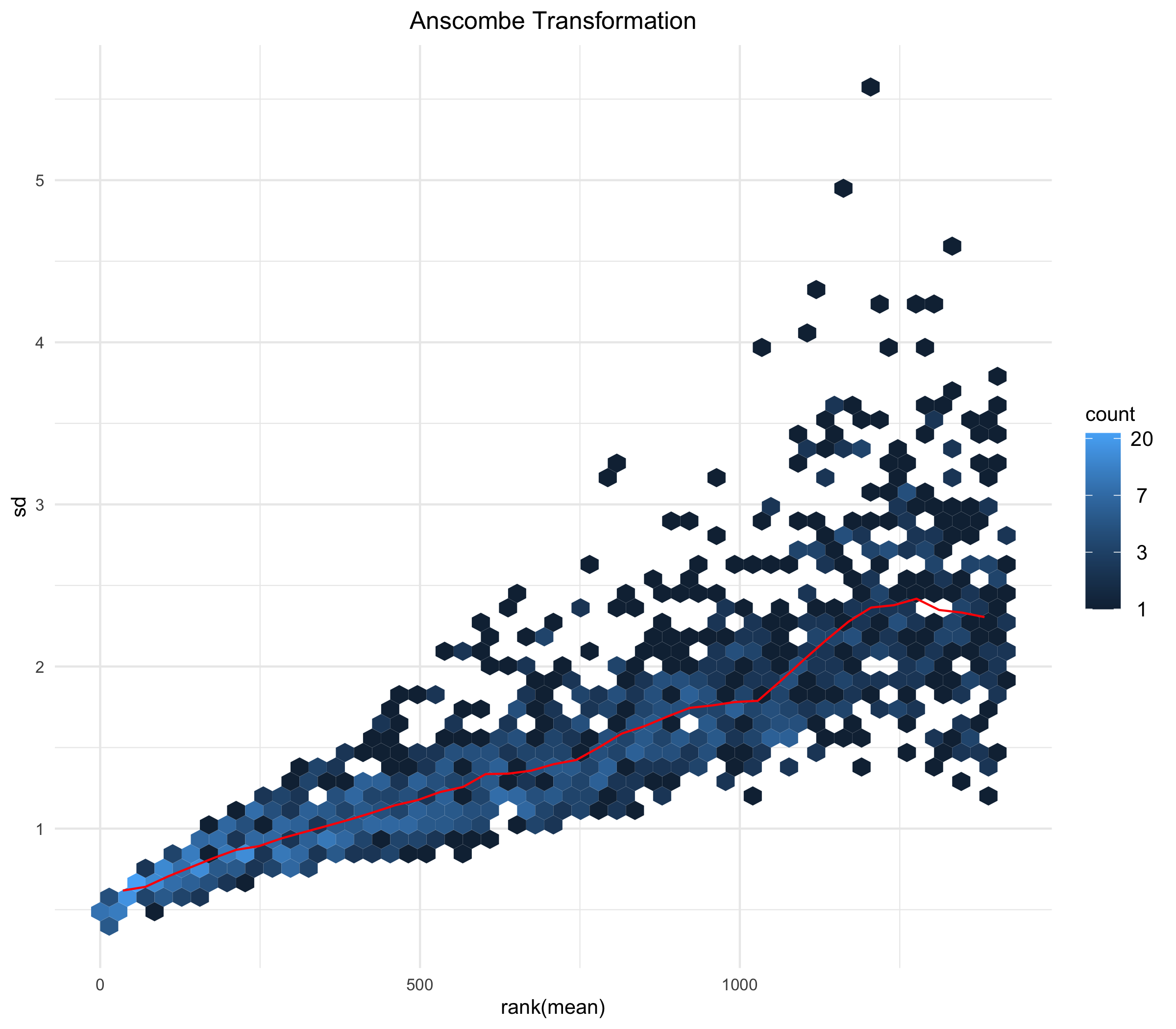}
		\caption{Anscombe's transformation of abundance data. Hexagonal binning avoids overplotting}
		\label{fig04}
	\end{figure}

	\begin{figure}[H]
		\centering
		\includegraphics[width=\linewidth]{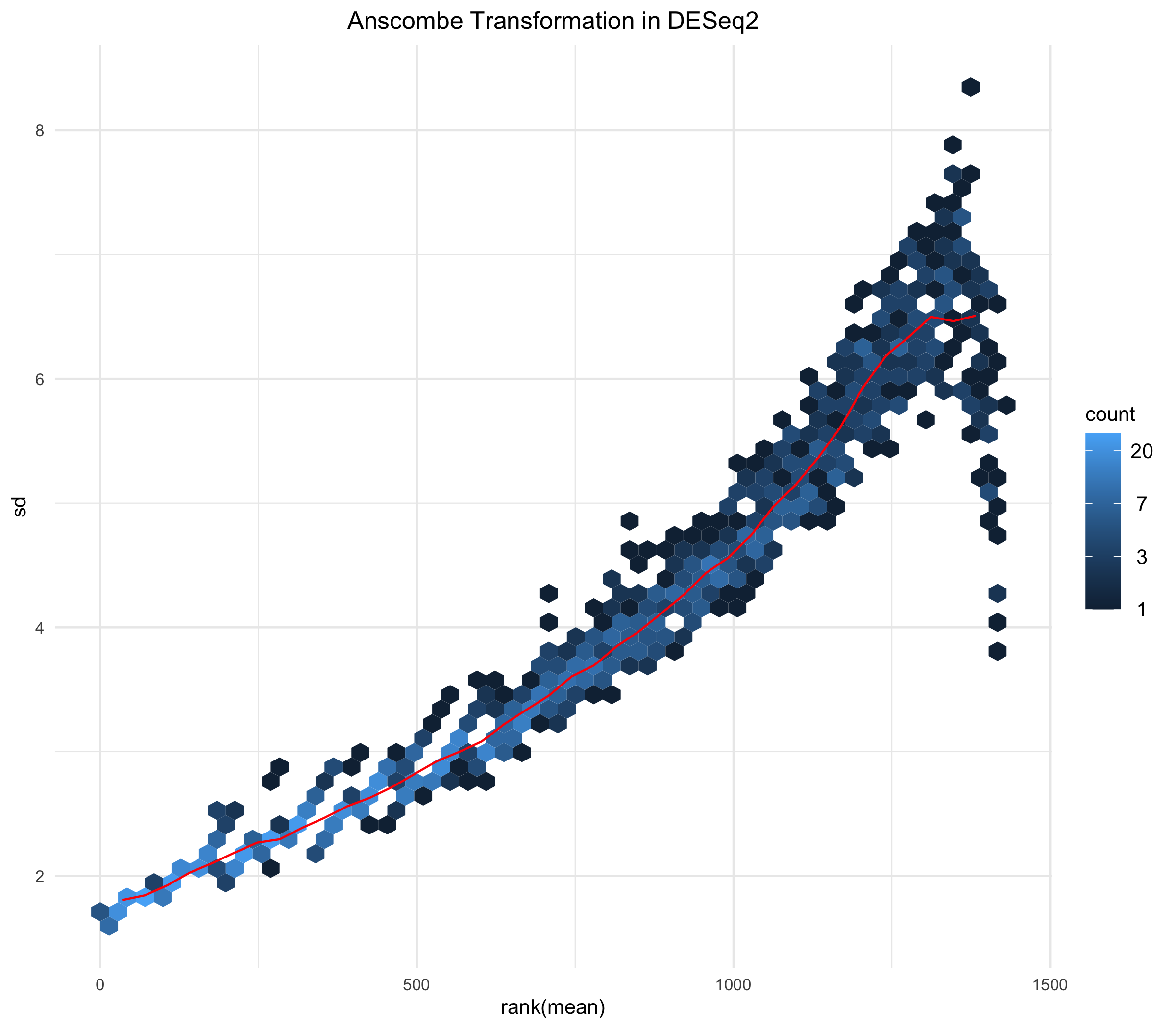}
		\caption{Anscombe's transformation implemented in {\tt DESeq2} package.}
		\label{fig06}
	\end{figure}

	\begin{figure}[H]
		\centering
		\includegraphics[width=\linewidth]{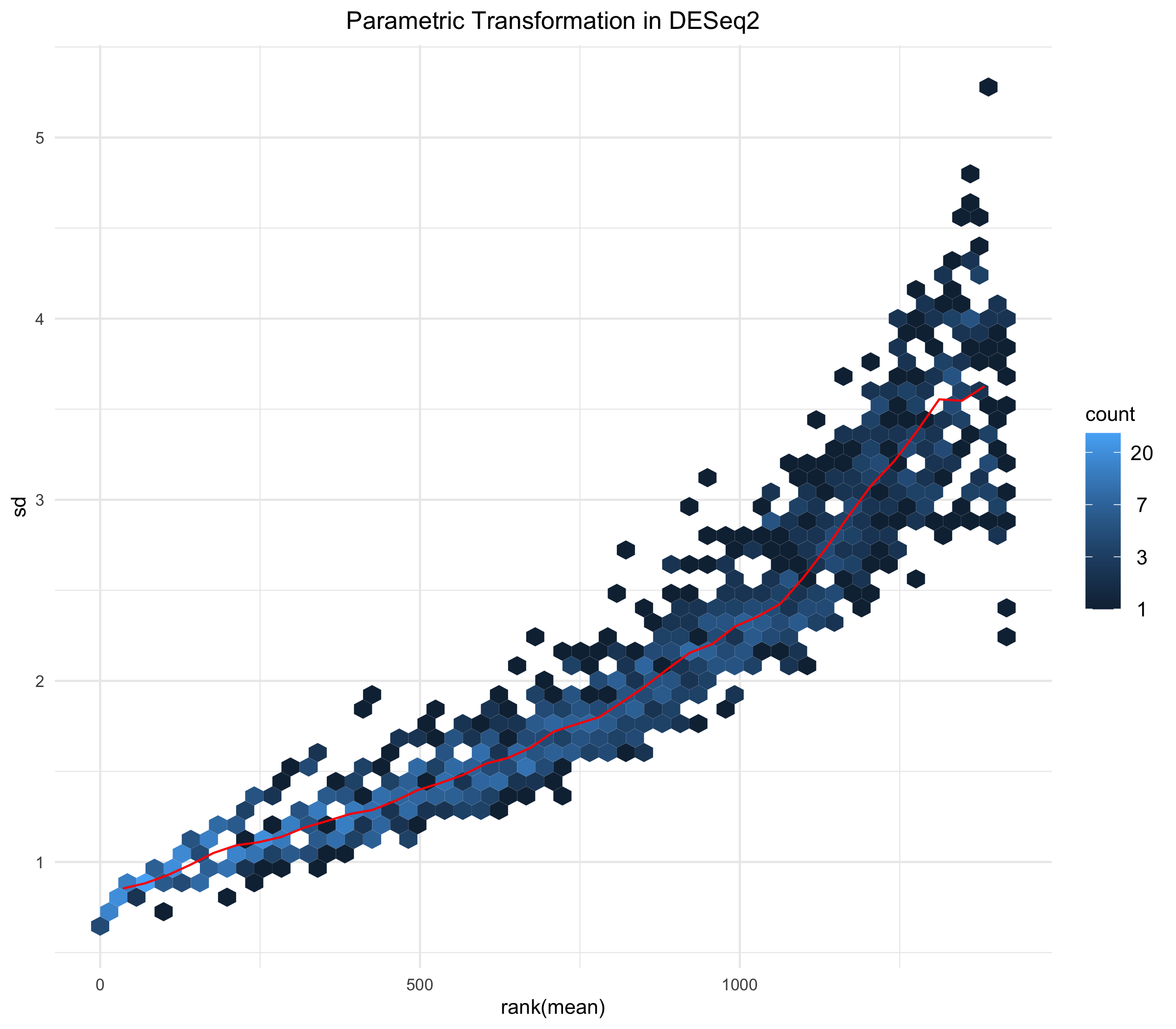}
		\caption{Parametric transformation implemented in {\tt DESeq2} package.}
		\label{fig07}
	\end{figure}

	\begin{figure}[H]
		\centering
		\includegraphics[width=\linewidth]{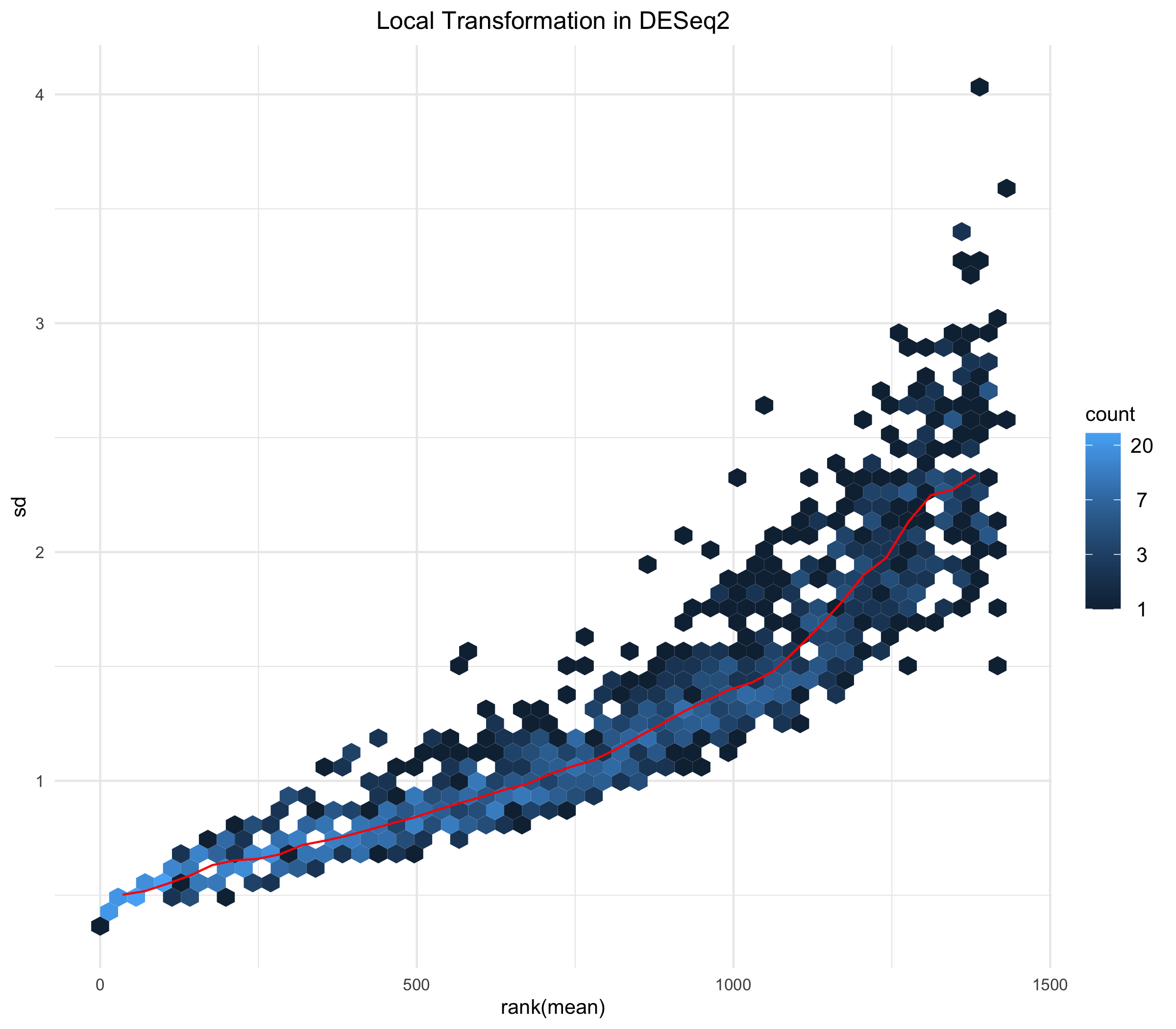}
		\caption{Nonparametric transformation implemented in {\tt DESeq2} package.}
		\label{fig08}
	\end{figure}

	\begin{figure}[H]
		\centering
		\includegraphics[width=\linewidth]{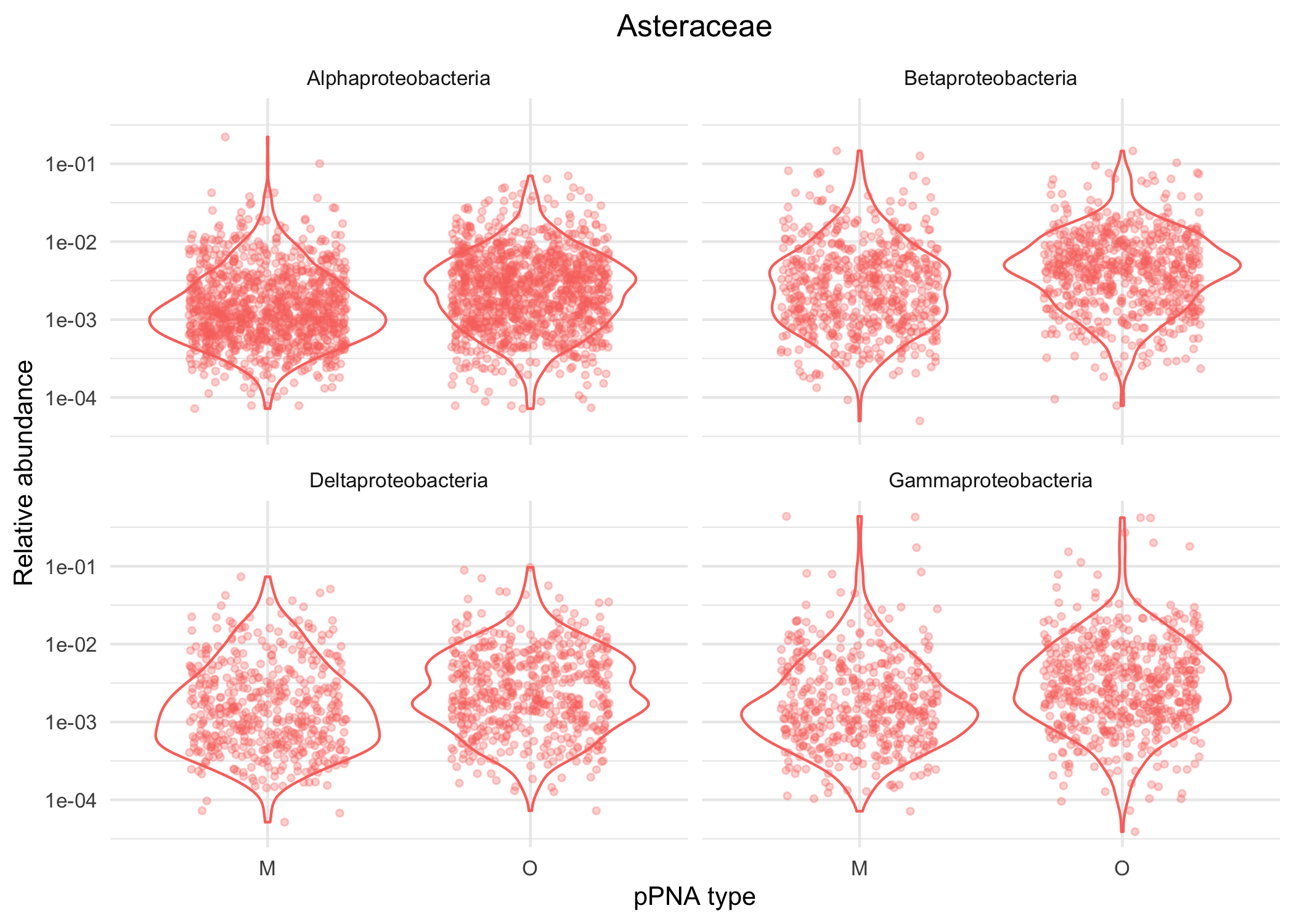}
		\caption{Distribution of relative abundance of four different Classes of \textit{Proteobacteria} in Asteraceae plants. O and M denote universal and Asteraceae-modified pPNA types, respectively.}
		\label{fig09}
	\end{figure}
	
	\begin{figure}[H]
		\centering
		\includegraphics[width=\linewidth]{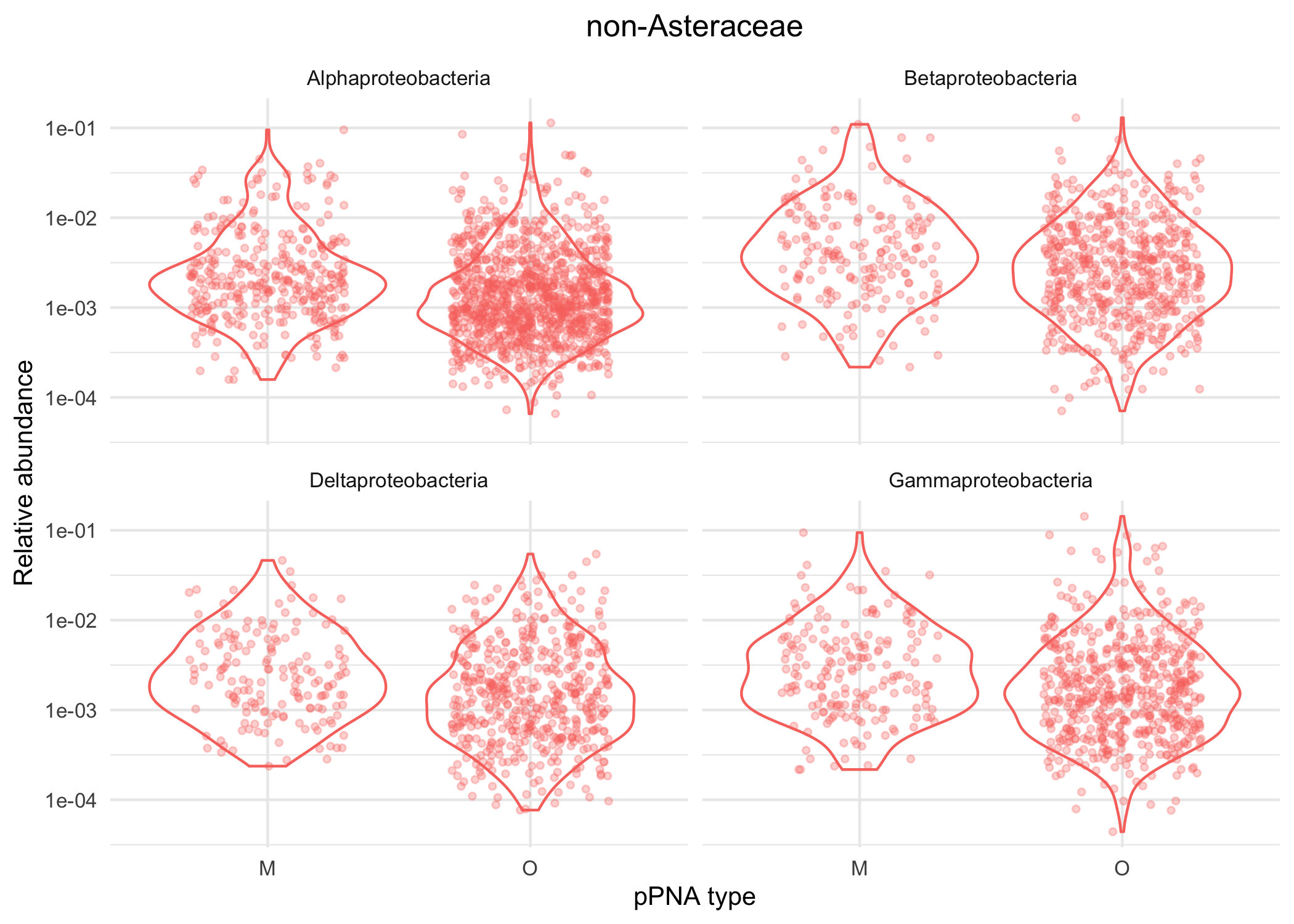}
		\caption{Distribution of relative abundance of four different Classes of \textit{Proteobacteria} in non-Asteraceae plants. O and M denote universal and Asteraceae-modified pPNA types, respectively. }
		\label{fig10}
	\end{figure}

	\begin{table}[H]
		\caption{Paired specimens in all plant types. }
		\label{tab04_paired}
		\centering
		\begin{tabular}{|l|l|l|l|}
			\hline
			Plant & Type & universal pPNA &  Asteraceae-modified pPNA \\
			\hline
			{\it Arctium minus} & Asteraceae&E-15-1 & E15-1\\
			{\it Sonchus arvensis} & Asteraceae &E-142-1 & E142-1\\
			{\it Sonchus arvensis} & Asteraceae &E-142-5 & E142-5\\
			{\it Sonchus arvensis} & Asteraceae &E-142-10 & E142-10\\
			{\it Sonchus oleraceus} & Asteraceae& E-143-2 & E143-2\\
			{\it Sonchus oleraceus} & Asteraceae& E-143-7 & E143-7\\
			{\it Centaurea solstitialis} & Asteraceae & ST-CAZ-4-R-O & ST-CAZ-4-R-M \\
			{\it Centaurea solstitialis} & Asteraceae & ST-SAL-22-R-O & ST-SAL-22-R-M \\
			{\it Centaurea solstitialis} & Asteraceae & ST-TRI-10-R-O & ST-TRI-10-R-M \\
			
			{\it Capsella bursa-pstoris} & non-Asteraceae &E-33-7 &E33-7\\
			{\it Capsella bursa-pstoris} & non-Asteraceae &E-33-8 &E33-8\\
			{\it Capsella bursa-pstoris} & non-Asteraceae &E-33-9&E33-9\\
			{\it Geum aleppicum} & non-Asteraceae &E-71-2 &E71-2\\
			{\it Geum aleppicum} & non-Asteraceae &E-71-3 &E71-3\\
			{\it Geum aleppicum} & non-Asteraceae &E-71-10 &E71-10\\
			{\it Phleum pratense} & non-Asteraceae &E-106-1 &E106-1\\
			{\it Phleum pratense} & non-Asteraceae &E-106-3 &E106-3\\
			{\it Phleum pratense} & non-Asteraceae &E-106-4 &E106-4\\
			\hline
		\end{tabular}
	\end{table}

	\begin{figure}[H]
		\centering
		\includegraphics[width=\linewidth]{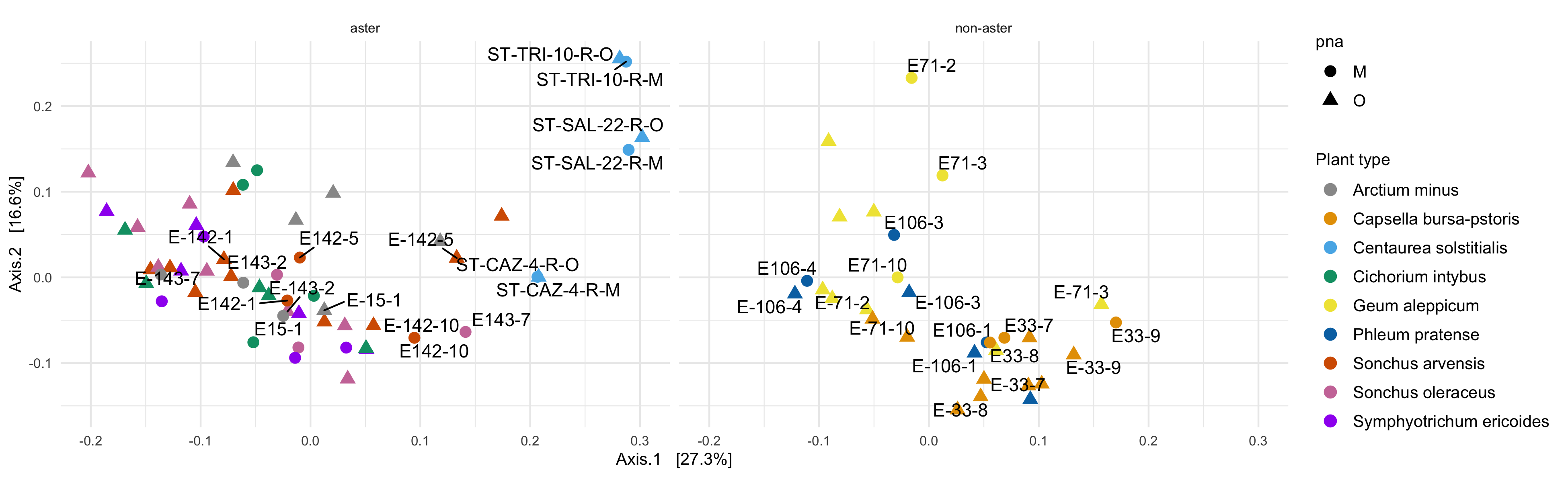}
		\caption{Multidimensional scaling (MDS) with weighted unifrac distance on all specimens. The shape denotes universal (O) and Asteraceae-modified (M) pPNA types, respectively. Facet denotes Asteraceae or non-Asteraceae plants. Paired specimens are labeled and make clusters, except paired-specimens (E-143-7, E143-7), (E-142-5, E142-5), (E-71-2, E71-2), and (E-71-3, E71-3) that are in positive and negative axes. {\it Centaurea solstitialis} specimens are outliers among Asteraceae plants in the positive direction of Axis 2. Axis 1 explains the microbial variability in plant types.}
		\label{mds_unifrac}
	\end{figure}
	
	\begin{figure}[H]
		\centering
		(A)
		\includegraphics[width=\linewidth]{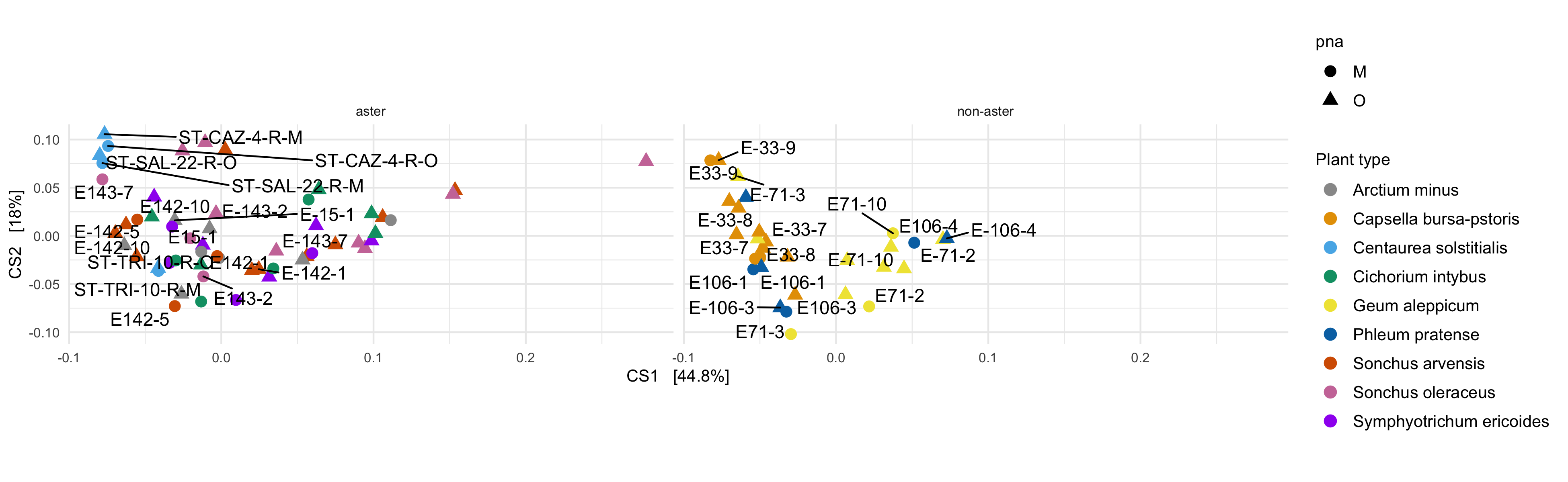}
		(B)
		\includegraphics[width=\linewidth]{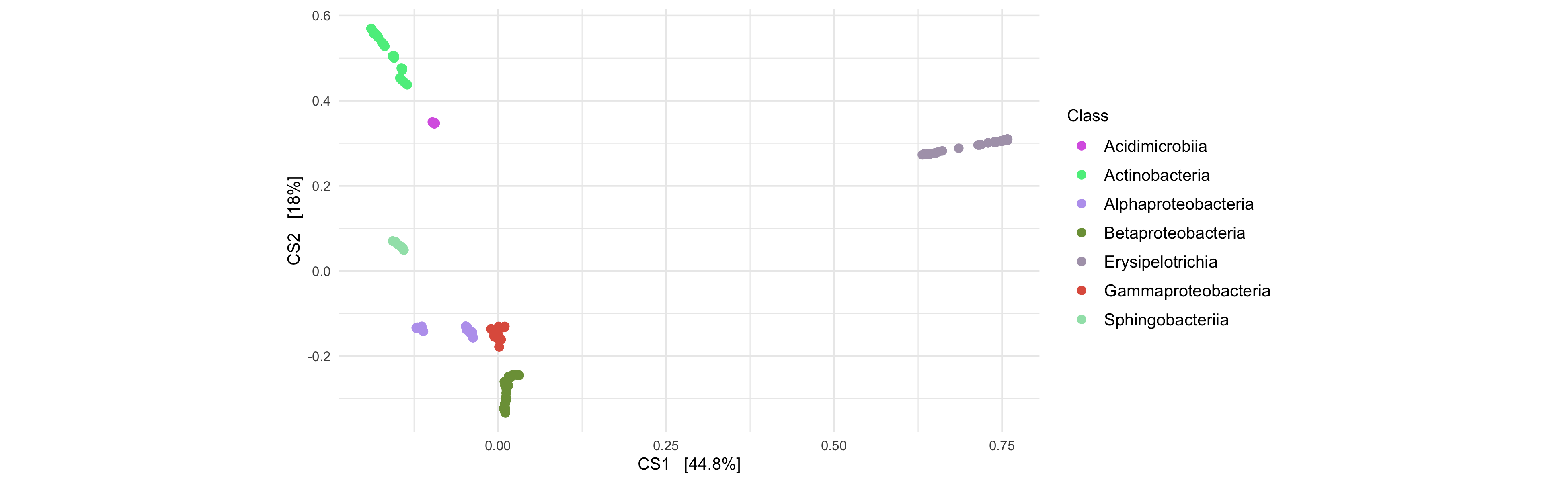}
		\caption{(A) A DPCoA plot that incorporates phylogenetic information. The shape denotes universal (O) and Asteraceae-modified (M) pPNA types, respectively. Facet denotes Asteraceae or non-Asteraceae plants. Paired specimens are labeled and make clusters, except paired-specimens (E-143-7, E143-7), (E-142-5, E142-5), and (E-71-3, E71-3) that are in positive and negative axes. Axis 1 explains the microbial variability in all paired specimens and specimens from {\it Sonchus oleraceus } and {\it Sonchus arvensis} plants with highly abundant {\it Erysipelotrichia}. (B) The DPCoA specimen ordination that is interpreted with respect to the ASV coordinates.}
		\label{fig1Vis}
	\end{figure}

	%%% network analysis
	\begin{figure}[H]
		\centering
		\includegraphics[width=\linewidth]{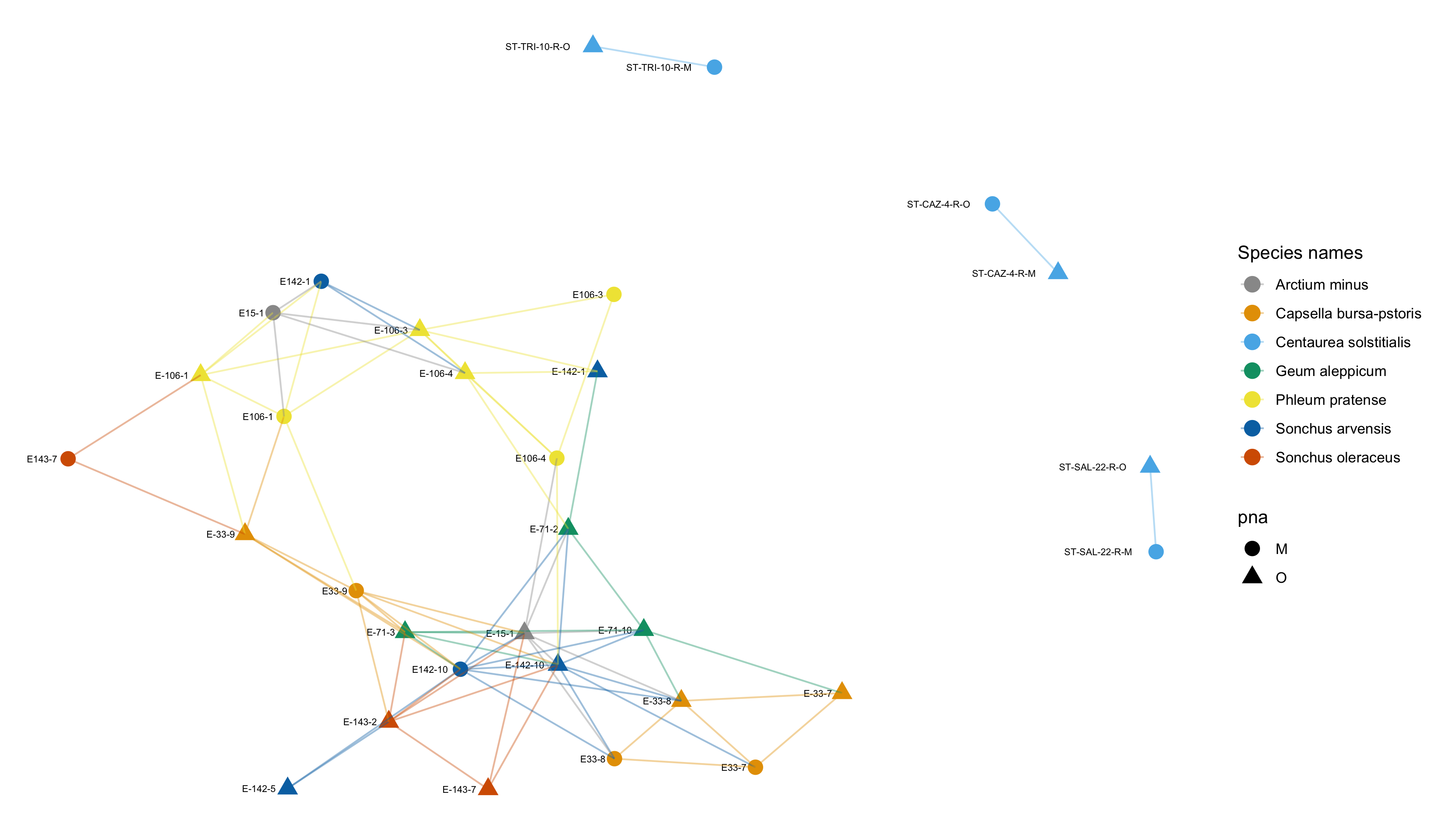}
		\caption{A network created by thresholding the Jaccard dissimilarity matrix at 0.8. All paired-specimens are connected, except (E-143-7, E143-7), (E-142-5, E142-5), (E-71-2, E71-2), and (E-71-3, E71-3).}
		\label{fig1net}
	\end{figure}

	\begin{figure}[H]
		\centering
		\includegraphics[width=\linewidth]{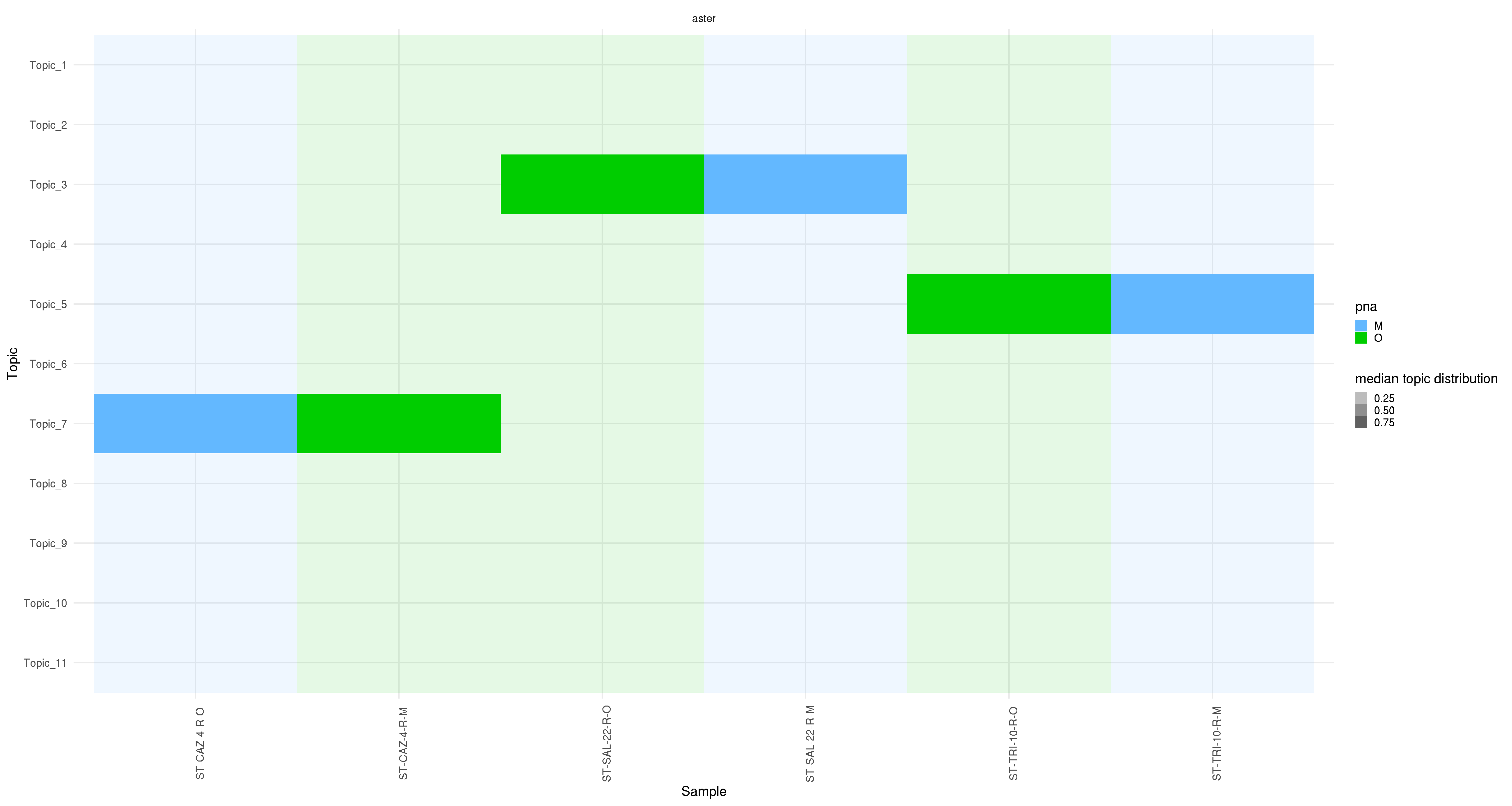}
		\caption{Topic distribution in specimens from \textit{Centaurea solstitialis} with eleven topics, which is from three different countries. This plant has three paired specimens sequenced with O and M-pPNA types. The color gradient represents the median topic distribution.}
		\label{fig_topic_centaurea_solsti_K_11}
	\end{figure}
	
	\begin{figure}[H]
		\centering
		\includegraphics[width=\linewidth]{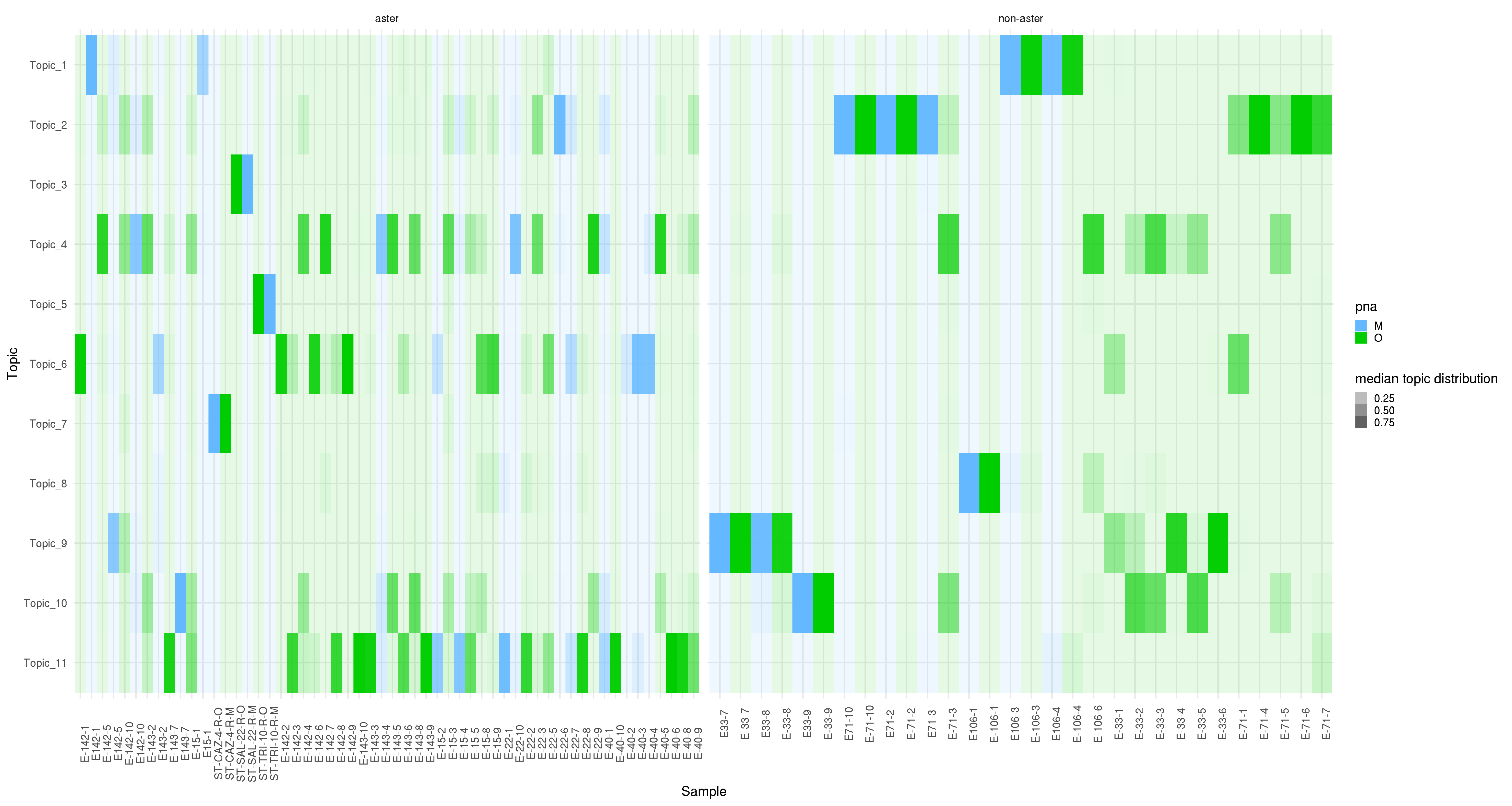}
		\caption{Topic distribution in all specimens with eleven topics. The color gradient represents the median topic distribution.}
		\label{fig_topic_dis_all_K_11}
	\end{figure}

	\begin{figure}[H]
		\centering
		\includegraphics[width=\linewidth]{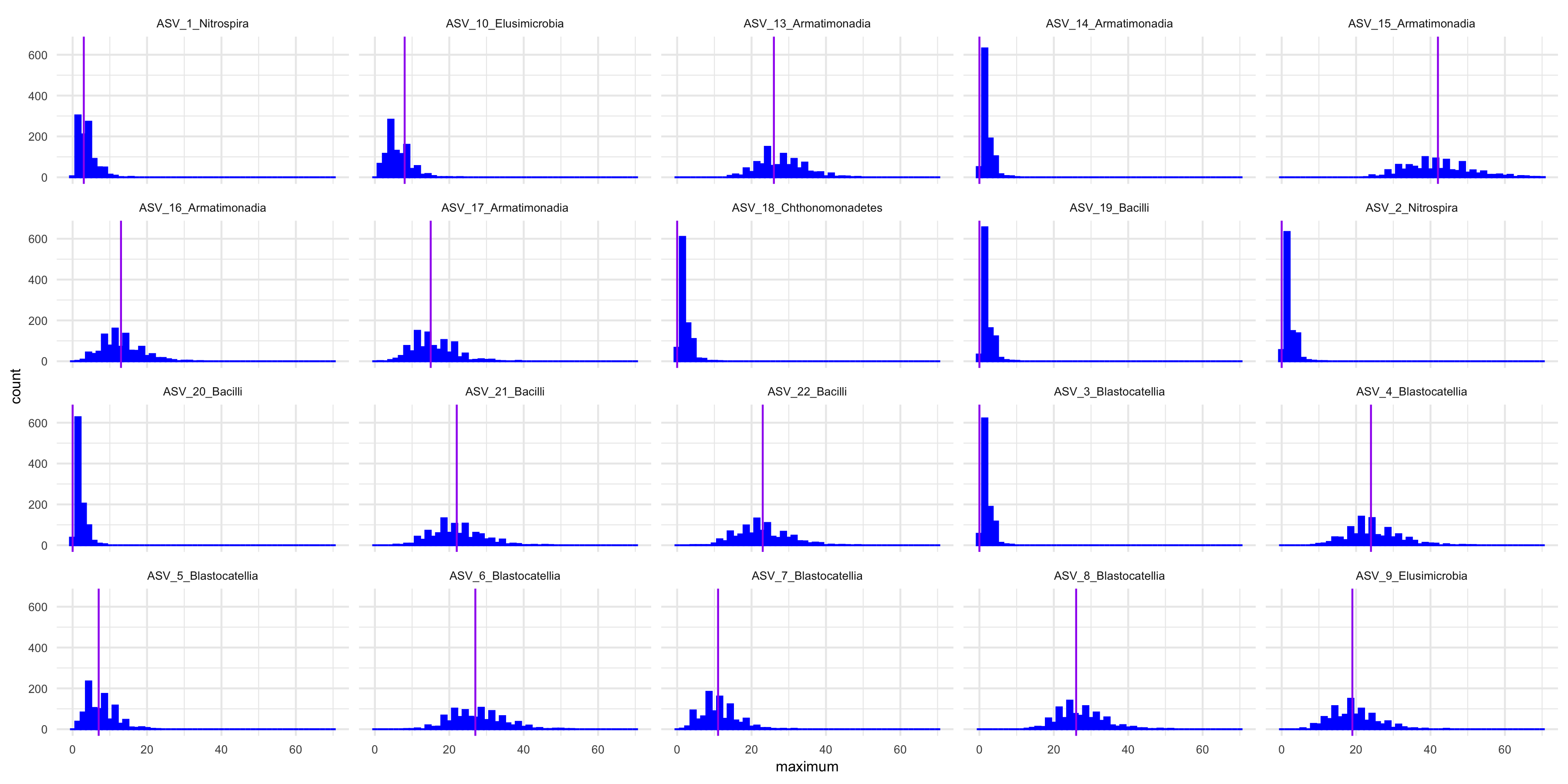}
		\caption{Predictive model check with simulated data, observed data, and a statistic $G\left(K_{ij} \right) = \text{max}\lbrace K_{ij} \rbrace$. Each facet shows the histogram of $G\left(K_{ij} \right)$ of each ASV in specimens from the posterior predictive distribution and the vertical line shows the value of $G\left(K_{ij} \right)$ of each ASV in observed data.}
		\label{fig_ppc_all_samples}
	\end{figure}
	
	\begin{figure}[H]
		\centering
		\includegraphics[width=\linewidth]{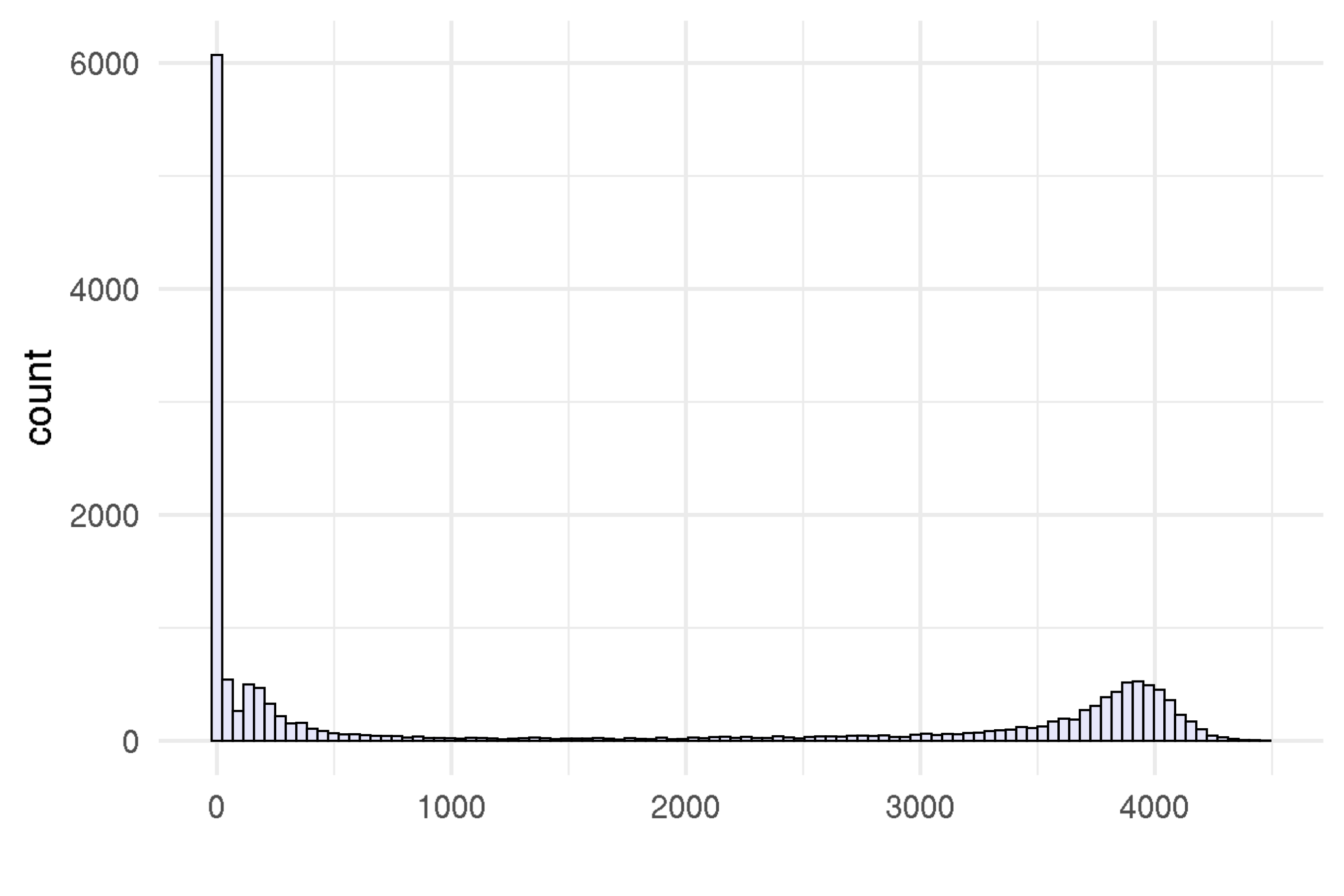}
		\caption{Effective sample size (ESS) with eleven topics.}
		\label{fig_ess_all_samples}
	\end{figure}

	\begin{figure}[H]
		\centering
		\includegraphics[width=\linewidth]{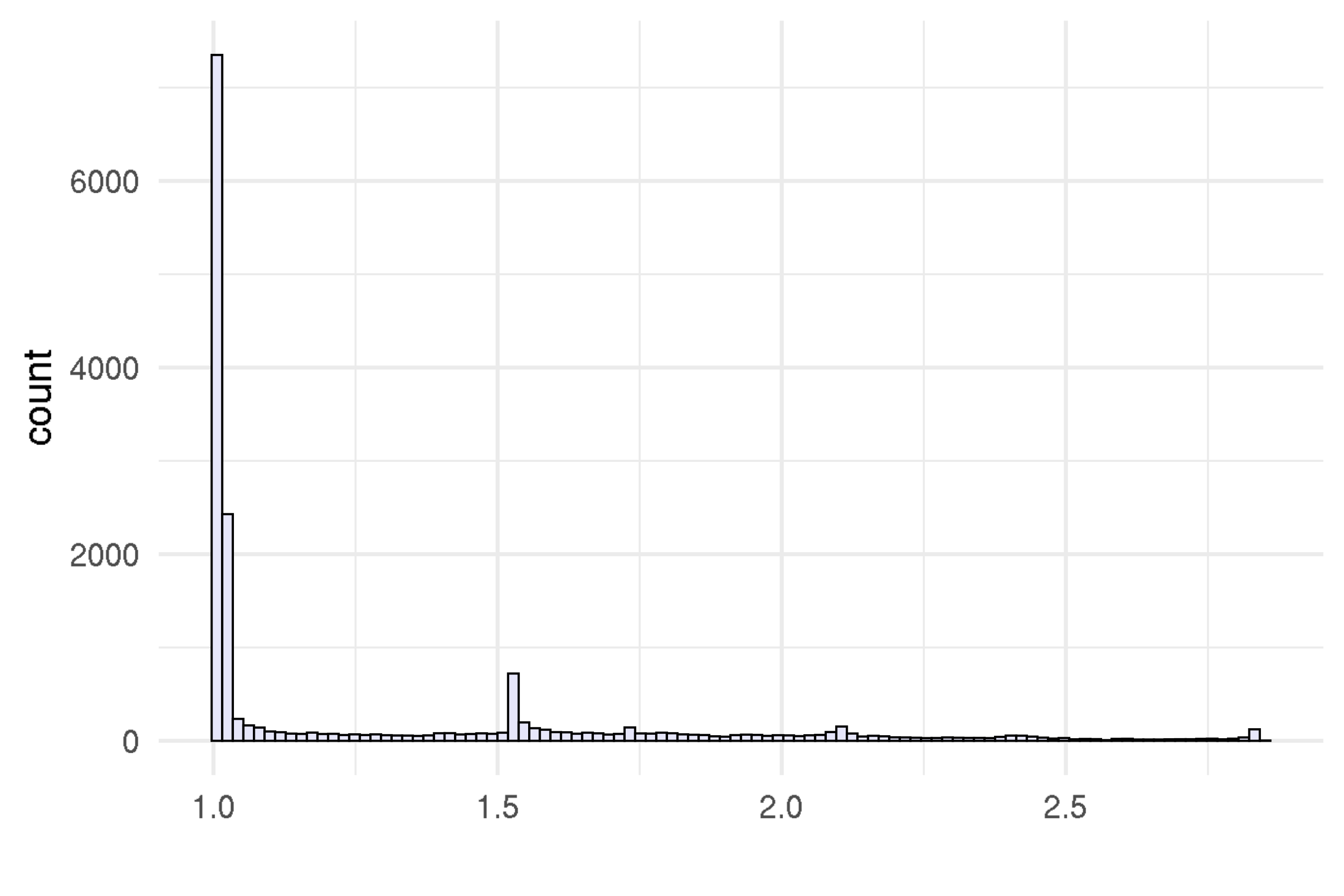}
		\caption{Split $\hat{R}$ with eleven topics.}
		\label{fig_rhat_all_samples}
	\end{figure}

	\begin{table}[H]
		\centering
		\caption{Generalized linear model results on median of topic proportion and covariate pPNA type on all specimens.}
		\label{tab_diffT_all}
		\begin{tabular}{rlrrrrl}
			\hline
			& Topic & lfc & lfcSE & WTS & pvalue & p.adj \\ 
			\hline
			1 & Topic\_1 & -0.90 & 0.95 & -0.95 & 0.34 & 0.6095 \\ 
			2 & Topic\_2 & 0.33 & 0.69 & 0.48 & 0.63 & 0.7196 \\ 
			3 & Topic\_3 & -0.56 & 0.58 & -0.96 & 0.34 & 0.6095 \\ 
			4 & Topic\_4 & 2.06 & 0.75 & 2.74 & 0.01 & 0.0224 \\ 
			5 & Topic\_5 & 2.69 & 0.67 & 4.05 & 0.00 & $<$.0001  \\ 
			6 & Topic\_6 & -0.29 & 0.80 & -0.36 & 0.72 & 0.7196 \\ 
			7 & Topic\_7 & -0.38 & 0.50 & -0.77 & 0.44 & 0.6095 \\ 
			8 & Topic\_8 & 0.64 & 0.75 & 0.85 & 0.39 & 0.6095 \\ 
			9 & Topic\_9 & 0.78 & 0.78 & 1.00 & 0.32 & 0.6095 \\ 
			10 & Topic\_10 & 2.62 & 0.73 & 3.61 & 0.00 & 0.0017 \\ 
			11 & Topic\_11 & 0.33 & 0.78 & 0.42 & 0.67 & 0.7196 \\ 
			\hline
		\end{tabular}
	\end{table}

	\begin{table}[H]
		\centering
		\caption{Generalized linear model results on median of topic proportion and covariate pPNA type on paired-specimens from Asteraceae plants.}
		\label{tab_diffT_all_aster_paired}
		\begin{tabular}{rlrrrrl}
			\hline
			& Topic & lfc & lfcSE & WTS & pvalue & p.adj \\ 
			\hline
			1 & Topic\_1 & -10.43 & 1.51 & -6.89 & 0.00 & $<$.0001 \\ 
			2 & Topic\_2 & 1.29 & 1.36 & 0.95 & 0.34 & 0.6245 \\ 
			3 & Topic\_3 & 0.58 & 1.41 & 0.41 & 0.68 & 0.8343 \\ 
			4 & Topic\_4 & 0.97 & 1.54 & 0.63 & 0.53 & 0.8281 \\ 
			5 & Topic\_5 & -0.15 & 1.28 & -0.12 & 0.91 & 0.9482 \\ 
			6 & Topic\_6 & 0.09 & 1.33 & 0.07 & 0.95 & 0.9482 \\ 
			7 & Topic\_7 & 0.45 & 1.03 & 0.44 & 0.66 & 0.8343 \\ 
			8 & Topic\_8 & -1.32 & 1.03 & -1.28 & 0.20 & 0.4524 \\ 
			9 & Topic\_9 & -7.81 & 1.33 & -5.89 & 0.00 & $<$.0001 \\ 
			10 & Topic\_10 & -1.53 & 1.21 & -1.27 & 0.21 & 0.4524 \\ 
			11 & Topic\_11 & 8.38 & 1.39 & 6.04 & 0.00 & $<$.0001 \\ 
			\hline
		\end{tabular}
	\end{table}

	\begin{table}[H]
		\centering
		\caption{Generalized linear model results on median of topic proportion and covariate pPNA type on paired- specimens  from non-Asteraceae plants.}
		\label{tab_diffT_all_nonaster_paired}
		\begin{tabular}{rlrrrrl}
			\hline
			& Topic & lfc & lfcSE & WTS & pvalue & p.adj \\ 
			\hline
			1 & Topic\_1 & -0.00 & 1.57 & -0.00 & 1.00 & 0.9985 \\ 
			2 & Topic\_2 & -0.24 & 1.49 & -0.16 & 0.87 & 0.9985 \\ 
			3 & Topic\_3 & -1.96 & 1.04 & -1.89 & 0.06 & 0.1633 \\ 
			4 & Topic\_4 & 4.33 & 1.20 & 3.59 & 0.00 & 0.0036 \\ 
			5 & Topic\_5 & 2.24 & 1.18 & 1.89 & 0.06 & 0.1633 \\ 
			6 & Topic\_6 & 0.10 & 0.88 & 0.11 & 0.91 & 0.9985 \\ 
			7 & Topic\_7 & -0.59 & 0.89 & -0.67 & 0.51 & 0.7945 \\ 
			8 & Topic\_8 & 0.63 & 1.51 & 0.42 & 0.68 & 0.9283 \\ 
			9 & Topic\_9 & -2.69 & 1.67 & -1.62 & 0.11 & 0.1941 \\ 
			10 & Topic\_10 & 2.42 & 1.50 & 1.62 & 0.11 & 0.1941 \\ 
			11 & Topic\_11 & 3.26 & 1.16 & 2.82 & 0.00 & 0.0267 \\ 
			\hline
		\end{tabular}
	\end{table}
	
\begin{supplement}	
\stitle{Code References}
\sdescription{Code to reproduce figures and simulations can be found at \href{https://pratheepaj.github.io/diffTop/ }{https://pratheepaj.github.io/diffTop/}.}	
\end{supplement}

\end{document}